\documentclass[conference]{formatting/IEEEtran}

\usepackage{graphicx}
\usepackage{multirow}
\usepackage{enumerate}
\usepackage{xspace}

\usepackage{listings}
\usepackage{float}
\makeatletter
\floatstyle{ruled}
\newfloat{algorithm}{tbp}{loa}
\providecommand{\algorithmname}{Algorithm}
\floatname{algorithm}{\protect\algorithmname}
\makeatother

\usepackage{shortcuts}

\usepackage[bookmarks=false]{hyperref}
\hypersetup{
    colorlinks = true,
    linkcolor = blue,
    anchorcolor = blue,
    citecolor = blue,
    filecolor = blue,
    urlcolor = blue
}

\renewcommand{\paragraph}{\vspace{3pt}\noindent\textbf}
\begin{document}

\date{}
\setlength{\parskip}{3.5pt}

\title{A Survey of \BCS}

\author{

\IEEEauthorblockN{Irfan Ul Haq}
\IEEEauthorblockA{IMDEA Software Institute \& \\
Universidad Polit\'ecnica de Madrid\\
irfanul.haq@imdea.org}

\and

\IEEEauthorblockN{Juan Caballero}
\IEEEauthorblockA{IMDEA Software Institute\\
juan.caballero@imdea.org}
}

\maketitle

\begin{abstract}

\noindent
\Bcs approaches compare two or more \chunks
to identify their similarities and differences.
The ability to compare binary code enables many real-world applications 
on scenarios where source code may not be available 
such as patch analysis, bug search, and malware detection and analysis. 
Over the past \numyears numerous \bcs approaches have been proposed, 
but the research area has not yet been systematically analyzed.
This paper presents a first survey of \bcs.
It analyzes \numapproaches \bcs approaches,
which are systematized on four aspects:
(1) the applications they enable,
(2) their approach characteristics,
(3) how the approaches are implemented, and
(4) the benchmarks and methodologies used to evaluate them.
In addition, the survey discusses the scope and origins of the area,
its evolution over the past two decades, and the challenges that lie ahead.

\end{abstract}


\section{Introduction}
\label{sec:intro}

\Bcs approaches compare two or more \chunks 
e.g., basic blocks, functions, or whole programs,
to identify their similarities and differences. 
Comparing binary code is fundamental in scenarios 
where the program source code is not available, which happens with 
commercial-of-the-shelf (COTS) programs, legacy programs, and malware.
\Bcs has a wide list of real-world applications such as
bug search~\cite{tracy,tedem,Pewny2015BugSearch,discovre,esh,
genius,bingo,binsequence,xmatch,gitz,vulseeker,gemini,firmup,binarm,alphadiff},
malware clustering~\cite{smit,mutantxs,Kim2016},
malware detection~\cite{Kruegel2005,Bruschi2006DSM,Cesare2014Control},
malware lineage~\cite{beagle,iline,Ming2015Memoized},
patch generation~\cite{exediff},
patch analysis~\cite{Flake2004Structural,Dullien2005Graph,
binhunt, mockingbird,binsequence,spain,Kargen2017},
porting information across program 
versions~\cite{bmat,Flake2004Structural,Dullien2005Graph}, and
software theft detection~\cite{cop}.

Identifying \bcs is challenging because 
much program semantics are lost due to the compilation process including
function names, variable names, source comments, and
data structure definitions.
Additionally, even when the program source code does not change, 
the binary code may change if the source is recompiled, 
due to secondary changes introduced by the compilation process. 
For example, the resulting binary code 
can significantly change when 
using different compilers, 
changing compiler optimizations, 
and selecting different target operating systems and CPU architectures.
Furthermore, obfuscation transformations can be applied on 
both the source code and the generated binary code, 
hiding the original code.

Given its applications and challenges, over the past \numyears 
numerous \bcs approaches have been proposed.
However, as far as we know there does not exist a systematic survey of this 
research area.
Previous surveys deal with 
binary code obfuscation techniques in packer tools~\cite{roundy2013binary}, 
binary code type inference~\cite{types}, and 
dynamic malware analysis techniques~\cite{egele2012survey}. 
Those topics are related because \bcs may need to tackle obfuscation, 
binary code type inference may leverage similar binary analysis platforms, 
and malware is often a target of \bcs approaches.
But, \bcs is well-separated from those topics 
as shown by previous surveys having 
no overlap with this paper on the set of papers analyzed.
Other surveys have explored similarity detection on any binary data, 
i.e., not specific to \emph{code}, 
such as hashing for similarity search~\cite{wang2014hashing} and 
similarity metrics on numerical and binary feature 
vectors~\cite{Cha2007, Choi2010}. 
In contrast, this survey focuses on approaches that compare 
binary code, i.e., that disassemble the executable byte stream. 

This paper presents a first survey of \bcs.
It first identifies \numapproaches \bcs approaches through a 
systematic selection process that examines over a hundred papers 
published in research venues 
from different computer science areas such as computer security, 
software engineering, programming languages, and machine learning. 
Then, it systematizes four aspects of those \numapproaches approaches:
(1) the applications they enable,
(2) their approach characteristics,
(3) how the approaches have been implemented, and
(4) the benchmarks and methodologies used to evaluate them.
In addition, it discusses the scope and origin of binary code similarity, 
its evolution over the past two decades, and the challenges that lie ahead.

\Bcs approaches widely vary in their approach, 
implementation, and evaluation. 
This survey systematizes each of those aspects, and 
summarizes the results in easy to access 
tables that compare the \numapproaches approaches across multiple dimensions, 
allowing beginners and experts 
to quickly understand their similarities and differences.
For example, the approach systematization includes, among others, 
the number of input \chunks being compared 
(e.g., one-to-one, one-to-many, many-to-many); 
the granularity of the \chunks analyzed 
(e.g., basic blocks, functions, programs); 
whether the comparison happens at the syntactical representation, 
the graph structure, or the code semantics;
the type of analysis used 
(e.g., static, dynamic, symbolic), and
the techniques used for scalability
(e.g., hashing, embedding, indexing).
The implementation systematization includes 
the binary analysis platforms used to build the approach,
the programming languages used to code it, 
the supported architectures for the input \chunks being compared, 
and whether the approach is publicly released.
The evaluation systematization covers the datasets on which the approaches 
are evaluated and the evaluation methodology including 
the type of evaluation (e.g., accuracy, comparison wih prior works, performance) and how the robustness of the approach is evaluated in face of 
common code transformations such as compiler and compilation option changes, 
different architectures, and obfuscation.

Beyond the systematization, 
this survey also discusses how \bcs has evolved from \diffing to 
\search and how the focus has moved from a single architecture to 
cross-architecture approaches. 
It shows that the present of the field is vibrant as many new approaches are 
still being proposed. 
It discusses technical challenges that remain open, but 
concludes that the future of the area is bright with important applications 
scenarios under way such as those related to \search engines and the 
Internet-of-Things. 

\paragraph{Paper structure.}
This paper is organized as follows. 
Section~\ref{sec:overview} provides an overview of \bcs. 
Section~\ref{sec:scope} details the scope of the survey and 
the paper selection process.
Section~\ref{sec:apps} summarizes applications of \bcs and
Section~\S\ref{sec:evolution} the evolution of the field over 
the last two decades.
Section~\ref{sec:approach} systematizes the characteristics of the 
\numapproaches \bcs approaches, 
Section~\ref{sec:implementation} their implementation, and 
Section~\ref{sec:evaluation} their evaluation. 
Finally, we discuss future research directions in Section~\ref{sec:discussion},
and conclude in Section~\ref{sec:conclusion}.
\section{Overview}
\label{sec:overview}

In this section, we first provide background on 
the compilation process (\S\ref{sec:compilation}). 
Then, we present an overview of the \bcs problem (\S\ref{sec:bcs}). 

\begin{figure*}[t]
\centering
\includegraphics[width=0.95\textwidth]{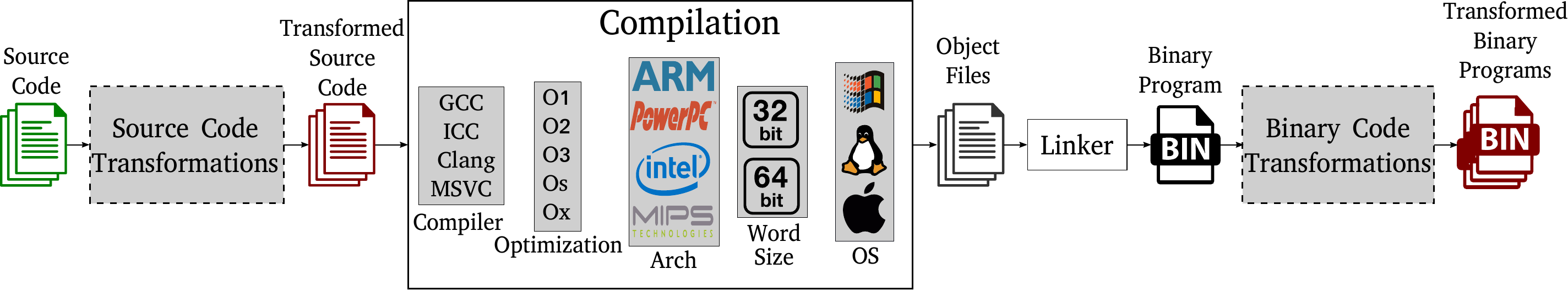}
\caption{
The extended compilation process. 
Dotted boxes are optional code transformations typically used for obfuscation. 
For a given source code, changing any of the grey boxes may produce a 
different binary program. 
}
\label{fig:compilation}
\end{figure*}

\subsection{Compilation Process}
\label{sec:compilation}

Binary code refers to the machine code that is produced by the
\emph{compilation process} and that can be run directly by a CPU.
The standard compilation process takes as input the 
source code files of a program. 
It compiles them with a chosen compiler and optimization level and 
for a specific platform (defined by the architecture, word size, and OS) 
producing object files. 
Those object files are then linked into a binary program, 
either a stand-alone executable or a library.

\Bcs approaches typically deal with an 
{\em extended compilation process}, illustrated in Figure~\ref{fig:compilation},
which adds two optional steps to the standard compilation process: 
source code and binary code transformations.
Both types of transformations are typically semantics-preserving 
(i.e., do not change the program functionality) and are 
most commonly used for obfuscation, 
i.e., to hamper reverse-engineering of the distributed binary programs.
Source code transformations happen pre-compilation.
Thus, their input and output are both source code.
They can be applied regardless of the target platform, but may be 
specific to the programming language used to write the program. 
On the other hand, binary code transformations happen post-compilation.
Thus, their input and output are binary code. 
They are independent of the programming language used, 
but may be specific to the target platform.

Obfuscation is a fundamental step in malware, but can also be applied 
to benign programs, e.g., to protect their intellectual property. 
There exist off-the-shelf obfuscation tools that use  
source code transformations (e.g., Tigress~\cite{tigress}), 
as well as binary code transformations (e.g., \emph{packers}~\cite{Ugarte14}).
Packing is a binary code transformation widely used by malware. 
Once a new version of a malware family is ready, the malware authors
pack the resulting executable to hide its functionality and
thus bypass detection by commercial malware detectors.  The packing process
takes as input an executable and produces another executable with the
same functionality, but with the original code hidden 
(e.g., encrypted as data and unpacked at runtime).
The packing process is typically applied many times
to the same input executable, creating polymorphic variants
of exactly the same source code,
which look different to malware detectors.
Nowadays, the majority of malware is packed
and malware often uses custom packers for which off-the-shelf unpackers
are not available~\cite{Ugarte14}.

A main challenge in \bcs is that the compilation process 
can produce different binary code representations for the same source code.
An author can modify any of the grey boxes in 
Figure~\ref{fig:compilation} to produce a different, 
but semantically-equivalent, binary program 
from the same source code.
Some of these modifications may be due to the standard compilation process. 
For example, to improve program efficiency an author may vary the 
compiler's optimization level, or change the compiler altogether. 
Both changes will transform the produced binary code, 
despite the source code remaining unchanged.
An author may also change the target platform to obtain a version of the 
program suitable for a different architecture.
In this case, the produced binary code may radically differ if the 
new target architecture uses a different instruction set.
An author may also deliberately apply obfuscation transformations to 
produce polymorphic variants of the same source code. 
The produced variants will typically have the same functionality defined by 
the original source code.
A desirable goal for \bcs approaches is that they are able to identify 
the similarity of binary code that corresponds to the same source code 
having undergone different transformations.
The \emph{robustness} of a \bcs approach captures
the compilation and obfuscation transformations that it can handle,
i.e., the transformations despite which it can still detect similarity.

\subsection{\BCS Overview}
\label{sec:bcs}

\Bcs approaches compare {\em \chunks}.
The three main characteristics of \bcs approaches are: 
(1) the type of the comparison
(identical, similar, equivalent), 
(2) the granularity of the \chunks being compared
(e.g., instructions, basic blocks, functions), and 
(3) the number of input \schunks being compared 
(one-to-one, one-to-many, many-to-many).
We detail these three characteristics next. 
For simplicity, we describe the comparison type and comparison granularity
for two inputs and then generalize to multiple inputs.

\paragraph{Comparison type.}
Two (or more) \chunks are identical if they have 
the same \emph{syntax}, i.e., the same representation.
The binary code can be represented in different ways such as 
an hexadecimal string of raw bytes, 
a sequence of disassembled instructions, or 
a control-flow graph.
Determining if several \chunks are identical is 
a Boolean decision (either they are identical or not) 
that it is easy to check: 
simply apply a cryptographic hash 
(e.g., SHA256) 
to the contents of each \schunk.
If the hash is the same, the \schunks are identical.
However, such straightforward approach fails to detect similarity 
in many cases.
For example, compiling the same program source code twice, 
using the same compilation parameters 
(i.e., same compiler version, same optimization level, 
same target platform)
produces two executables with different file hash.
This happens because the executable may include metadata
that differs in both compilations such as the
compilation date, which is automatically computed and included into the 
header of the generated executable.

Two \chunks are equivalent if they have the 
same \emph{semantics}, 
i.e., if they offer exactly the same functionality.
Equivalence does not care about the syntax of the binary code. 
Clearly, two identical \chunks will have the same semantics, 
but different \chunks may as well.
For example,  \texttt{mov \%eax,\$0} and \texttt{xor \%eax,\%eax} 
are semantically equivalent x86 instructions 
because both set the value of register \texttt{EAX} to zero.
Similarly, the same source code compiled for two different target 
architectures should produce equivalent executables, whose syntax 
may be completely different if the architectures use different instruction sets.
Proving that two arbitrary programs are functionally equivalent is an 
undecidable problem that reduces to solving the halting problem~\cite{binjuice}.
In practice, determining binary code equivalence is a very 
expensive process that can only be performed for small \chunks. 

Two \chunks can be considered similar if their 
syntax, structure, or semantics are similar.
Syntactic similarity compares the code representation. 
For example, clone detection approaches consider that a target \chunk 
is a clone of some source binary code if their syntax are similar. 
Structural similarity compares graph representations of binary code 
(e.g., control flow graphs, call-graphs). 
It sits between syntactic and semantic similarity. 
The intuition is that the control flow of the binary code captures 
to some extent its semantics, e.g., the decisions taken on the data. 
Furthermore, the graph representation captures 
multiple syntactic representations of the same functionality.
However, it is possible to modify the graph structure without affecting 
the semantics, e.g., by inlining functions.
Semantic similarity compares the code functionality. 
A simple approach to semantic similarity compares 
the interaction between the program and its environment through OS APIs or 
system calls.
But, two programs with similar system calls can perform significantly 
different processing on their output, 
so more fine-grained semantic similarity approaches focus on a 
syntax-independent comparison of the code.

Generally speaking, the more robust an approach is, 
i.e., the more transformations it can capture,
the more expensive it also is. 
Syntactic similarity approaches are cheapest to compute, 
but least robust.
They are sensitive to simple changes in the binary code, 
e.g., register reallocation, instruction reordering, 
replacing instructions with semantically equivalent ones.
Structural similarity sits in the middle. 
It is robust against multiple syntactical transformations, 
but sensitive to transformations that change code structure such as 
code inlining or removal of unused function parameters.
Semantic similarity is robust against semantics-preserving transformations, 
despite changes to the code syntax and structure, 
but it is very expensive to compute for large \chunks. 

\begin{figure}[t]
\centering
\includegraphics[width=0.4\textwidth]{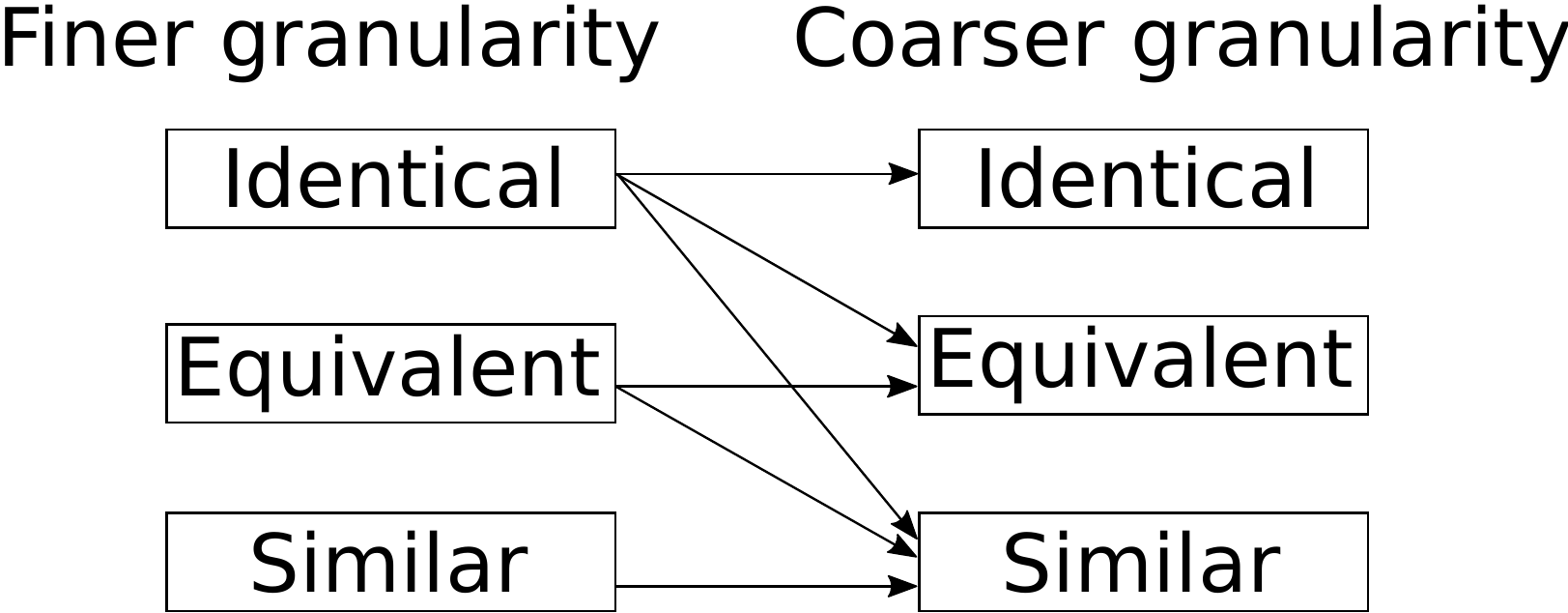}
\caption{Similarity at a finer granularity can be used to infer 
a different type of similarity at a coarser granularity.
}
\label{fig:granularity}
\end{figure}

\paragraph{Comparison granularity.}
\Bcs approaches can be applied at different granularities. 
Common granularities are 
instructions; 
basic blocks; 
functions; and
whole programs. 
To perform a comparison at a coarser granularity, 
some approaches use a different comparison at a finer granularity, 
and then combine the finer granularity results.
For example, to compare whether two programs are similar, 
an approach could determine the fraction of identical functions 
between both programs.
Thus, we differentiate between the \emph{input granularity}, 
i.e., the granularity of the input \chunks that the approach compares,
and the \emph{approach granularities}, 
i.e., the granularities of the different comparisons in the approach.

Applying a specific comparison at a finer granularity may restrict the type of 
comparison that can be performed at a coarser granularity, 
as illustrated in Figure~\ref{fig:granularity}.
The figure shows that computing whether two \chunks are 
identical at a finer granularity (e.g., basic block) can be used to 
compute that the coarser granularity \schunks that encompass them 
(e.g., their functions) are identical, equivalent, or similar. 
However, similarity at a finer granularity cannot be used to  
infer that the coarser granularity code is equivalent or identical.
For example, when comparing two functions, 
just because all their basic blocks are similar, 
it cannot be concluded that the functions are identical or equivalent.
On the other hand, similarity is the most general type of comparison 
and any finer granularity comparison type can be used to infer it.

\paragraph{Number of inputs.}
\Bcs approaches can compare two or more \chunks. 
Those that compare more than two \schunks 
can further be split into comparing one \schunk to the rest or 
comparing each \schunk to all other \schunks.
Thus, we identify three types of approaches based on the number of inputs and 
how they are compared: 
one-to-one ($OO$), one-to-many ($OM$), and many-to-many ($MM$). 
The source of the input \schunks is application-dependent.
They may come 
from the same program version 
(e.g., two functions of the same executable), 
from two versions of the same program, 
and from two different programs.

One-to-one approaches compare an {\em original} \chunk 
(also called source, old, plaintiff, or reference) to a {\em target} \chunk
(also called new, patched, or upgrade).
Most $OO$ approaches perform {\em \diffing}, 
i.e., they diff two consecutive, or close, versions of the 
same program to identify what was added, removed, or modified in the 
target (patched) version.
The granularity of \diffing is most often functions and the 
diffing tries to obtain a mapping between a function in the original 
program version and another function in the target program version. 
Added functions are original functions that cannot be mapped to a 
target function; 
removed functions are target functions that cannot be mapped to an 
original function; and 
modified functions are mapped functions that are not identical.

One-to-many approaches compare a {\em query} \chunk to many target \chunks. 
Most $OM$ approaches perform {\em \search}, 
i.e., they search if the query \schunk is 
similar to any of the target \schunks and 
return the top $k$ most similar target \chunks.
The target \schunks may come from multiple versions of the same program 
(different than the version the query \schunk comes from), 
from different programs compiled for the same architecture, or 
from programs compiled for different architectures.

In contrast to $OO$ and $OM$ approaches, 
many-to-many approaches do not distinguish between source and target \schunks.
All input \schunks are considered equal and compared against each other. 
These approaches typically perform {\em \clustering}, 
i.e., they output groups of similar \chunks called clusters.

\section{Scope \& Paper Selection}
\label{sec:scope}

To keep our survey of the state-of-the-art focused and manageable 
it is important to define what is, and what is not, within scope.
Overall, the main restriction is that we focus on works that compare 
binary code.
This restriction, in turn, introduces the following four constraints:

\begin{enumerate}

\item We exclude approaches that require access to the source code, 
namely source-to-source (e.g.,~\cite{ccfinder}) and 
source-to-binary (e.g.,~\cite{fiber}) similarity approaches.

\item We exclude approaches that operate on bytecode 
(e.g.,~\cite{andarwin,3d-cfg}).

\item We exclude behavioral approaches that compare similarity exclusively
on the interaction of a program with its environment through
system calls or OS API calls 
(e.g.,~\cite{forrest1996sense,kirda2006behavior,minimal,Bayer2009Scalable}).

\item We exclude approaches that consider binary code as a sequence of 
raw bytes with no structure such as 
file hashes (e.g., \cite{pehash}), 
fuzzy hashes (e.g., \cite{ssdeep,tlsh}), and 
signature-based approaches (e.g.,~\cite{polygraph,hamsa}).
Approaches need to disassemble raw bytes into instructions to be considered.
While we do not include the papers describing byte-level approaches, 
we do examine the use of some of those techniques (e.g., fuzzy hashing) 
by the analyzed approaches.

\end{enumerate}

In addition, we introduce the following constraints to keep the 
scope of the survey manageable:

\begin{enumerate}

\setcounter{enumi}{4}

\item We limit the survey to papers published on peer-reviewed venues
      and technical reports from academic institutions.
      Thus, we do not analyze tools, 
      but rather the research works describing their approach 
      (e.g., \cite{Flake2004Structural,Dullien2005Graph} for \bindiff).

\item We exclude papers that do not propose a new \bcs approach or technique, 
      but simply apply off-the-shelf \bcs tools as a step 
      towards their goal.

\end{enumerate}

\paragraph{Paper selection.}
To identify candidate papers, we first systematically 
examined all papers published in the last \numyears in 14 top venues 
for computer security and software engineering: 
\emph{IEEE S\&P},       \emph{ACM CCS},         \emph{USENIX Security}, \emph{NDSS},            \emph{ACSAC},           \emph{RAID},            \emph{ESORICS},         \emph{ASIACCS},         \emph{DIMVA},           \emph{ICSE},            \emph{FSE},             \emph{ISSTA},           \emph{ASE}, and         \emph{MSR}.             Not all relevant \bcs approaches have been published in those venues, 
which is especially true for early approaches. 
To identify candidate papers in other venues, 
we extensively queried specialized search engines such as Google Scholar
using terms related to \bcs and its applications, 
e.g., code search, binary diffing, bug search.
We also carefully examined the references of the candidate papers 
for any further papers we may have missed. 
This exploration identified over a hundred candidate papers. 
We then read each candidate paper to determine if they 
proposed a \bcs approach that satisfied the above scope constraints.

In the end, we identified the \numapproaches \bcs 
research works in Table~\ref{tab:approach}, 
whose approaches are systematized. 
The first three columns of Table~\ref{tab:approach} capture
the name of the approach, 
the year of publication, and 
the venue where the work was published.
The research works are sorted by publication date
creating a timeline of the development of the field.
Papers are identified by their system name, if available, otherwise
by the initials of each author's last name and the year of publication. 
The \numapproaches papers have been published in \numvenues venues. 
\Bcs is quite multidisciplinary;
while most papers appear in computer security venues (36 papers in 20 venues), 
there are works in software engineering (13 papers in 8 venues), 
systems (6 papers in 4 venues), and 
machine learning (2 papers in 2 venues).
The venues with most \bcs papers are:
DIMVA~(6), ASE~(4), CCS~(3), USENIX Security~(3), and PLDI~(3).

\begin{table*}
\centering
\scriptsize
\caption{Comparison among \bcs approaches.
For boolean columns \Y~means supported and \N~unsupported.
Input Comparison can be one-to-one (OO), one-to-many (OM),
or many-to-many (MM). 
Input Granularity and Approach Granularities can be 
instruction (I),
basic block (B),
function (F), 
or program (P).
Approach Comparison can be Similar (S), Identical (I), or 
Equivalent (E).
Structural similarity can use CFG (C), ICFG (I), Callgraph (G), and 
other custom graphs (O).
Machine learning can be Supervised (S) or Unsupervised (U).
In Normalization, 
\N~means no normalization,
\D~operand removal,
\B~operand normalization,
\SC~mnemonic normalization, and
\SR~code elimination.
\label{tab:approach}}
{
\renewcommand\featuretext[1]{\rotatebox{90}{#1}}
\begin{tabular}{|r|l|l|cllllcccccc|ccc|c|} 
\hline
& & & \multicolumn{15}{c|}{\textbf{Approach Characteristics}} \\
\cline{4-18}

& & & & & & & & & & & & & & & & & \\

\textbf{\system} & \textbf{Year} & \textbf{Venue} &
\featuretext{\textbf{Input Comparison}} &
\featuretext{\textbf{Approach Comparison}} &
\featuretext{\textbf{Input Granularity}} & \featuretext{\textbf{Approach Granularities}} & \featuretext{\textbf{Syntactical similarity}} & 
\featuretext{\textbf{Semantic similarity}} & 
\featuretext{\textbf{Structural similarity}} & 
\featuretext{\textbf{Feature-based}} &
\featuretext{\textbf{Machine learning}} &
\featuretext{\textbf{Locality sensitive hashing}} &

\featuretext{\textbf{Cross-architecture}} & 

\featuretext{\textbf{Static analysis}} &
\featuretext{\textbf{Dynamic analysis}}  & 
\featuretext{\textbf{Dataflow analysis}}  & 

\featuretext{\textbf{Normalization}}

\\ \hline

\exediff~\cite{exediff}                 & 1999 & WCSSS      & OO  & I   & P  & I     & \Y & \N & \N  & \N & \N & \N & \N &    \Y & \N & \N &   \B        \\
\bmat~\cite{bmat}                       & 1999 & FDO2       & OO  & S,I & P  & F,B   & \Y & \N & C   & \N & \N & \N & \N &    \Y & \N & \N &   \D\B\SC   \\ 
\flake~\cite{Flake2004Structural}       & 2004 & DIMVA      & OO  & S   & P  & F     & \N & \N & C,G & \Y & \N & \N & \N &    \Y & \N & \N &   \N        \\ 
\dullien~\cite{Dullien2005Graph}        & 2005 & SSTIC      & OO  & S,I & P  & F,B,I & \Y & \N & C,G & \Y & \N & \N & \N &    \Y & \N & \N &   \D\B      \\ 
\kruegel~\cite{Kruegel2005}             & 2005 & RAID       & MM  & S   & P  & B*    & \N & \Y & I   & \N & \N & \N & \N &    \Y & \N & \N &   \D\B      \\ 
\bruschi~\cite{Bruschi2006DSM}          & 2006 & DIMVA      & OO  & S   & P  & B*    & \N & \Y & I   & \N & \N & \N & \N &    \Y & \N & \Y &   \D\B      \\ 
\binhunt~\cite{binhunt}                 & 2008 & ICISC      & OO  & S,E & P  & F,B   & \N & \Y & C,G & \N & \N & \N & \N &    \Y & \N & \Y &   \N        \\ 
\saebjornsen~\cite{Saebjornsen2009DCC}  & 2009 & ISSTA      & MM  & S,I & I* & I*    & \Y & \N & \N  & \Y & \N & \Y & \N &    \Y & \N & \N &   \B        \\ 
\smit~\cite{smit}                       & 2009 & CCS        & OM  & S,I & P  & F     & \Y & \N & G   & \Y & \N & \N & \N &    \Y & \N & \N &   \N        \\ 
\idea~\cite{idea}                       & 2010 & ESSoS      & MM  & S   & P  & I*    & \Y & \N & \N  & \Y & \N & \N & \N &    \Y & \N & \N &   \B        \\ 
\mbc~\cite{mbc}                         & 2012 & RACS       & MM  & S   & P  & B     & \Y & \N & \N  & \Y & \N & \N & \N &    \Y & \N & \N &   \B        \\ 
\ibinhunt~\cite{ibinhunt}               & 2012 & ICISC      & OO  & S,E & P  & B     & \N & \Y & I   & \N & \N & \N & \N &    \Y & \Y & \Y &   \N        \\ 
\beagle~\cite{beagle}                   & 2012 & ACSAC      & MM  & S   & P  & B*    & \N & \Y & C   & \N & \N & \N & \N &    \Y & \Y & \N &   \B        \\ 
\binhash~\cite{binhash}                 & 2012 & ICMLA      & MM  & E   & F  & B     & \N & \Y & \N  & \Y & U  & \Y & \N &    \Y & \N & \Y &   \B        \\ 
\binjuice~\cite{binjuice}               & 2013 & PPREW      & OO  & S,E & P  & F,B   & \N & \Y & \N  & \N & \N & \N & \N &    \Y & \N & \Y &   \N        \\
\binslayer~\cite{binslayer}             & 2013 & PPREW      & OO  & S   & P  & F,B   & \N & \N & C,G & \N & \N & \N & \N &    \Y & \N & \N &   \D\B      \\ 
\rendezvous~\cite{rendezvous}           & 2013 & MSR        & OM  & S   & F  & F     & \Y & \N & \N  & \N & \N & \N & \N &    \Y & \N & \N &   \D\B      \\ 
\mutantxs~\cite{mutantxs}               & 2013 & Usenix ATC & MM  & S   & P  & I*    & \Y & \N & \N  & \Y & U  & \N & \N &    \Y & \N & \N &   \D\B        \\ 
\expose~\cite{expose}                   & 2013 & COMPSAC    & OM  & S,E & P  & F,I*  & \Y & \Y & \N  & \Y & \N & \N & \N &    \Y & \N & \Y &   \B        \\ 
\iline~\cite{iline}                     & 2013 & USENIX Sec & MM  & S   & P  & B,I*  & \Y & \N & \N  & \Y & U  & \N & \N &    \Y & \Y & \N &   \B\SC\SR  \\ 
\lee~\cite{Lee2013}                     & 2013 & RACS       & OO  & S   & P  & F,I*  & \N & \N & C,G & \Y & \N & \N & \N &    \Y & \N & \N &   \D\B      \\ 
\tracy~\cite{tracy}                     & 2014 & PLDI       & OM  & S,E & F  & I*    & \Y & \Y & \N  & \N & \N & \N & \N &    \Y & \N & \Y &   \SR       \\ 
\binclone~\cite{binclone}               & 2014 & SERE       & MM  & S,I & I* & I*    & \Y & \N & \N  & \Y & \N & \N & \N &    \Y & \N & \N &   \B        \\ 
\ruttenberg~\cite{Ruttenberg2014Shared} & 2014 & DIMVA      & MM  & S   & F* & F     & \N & \N & \N  & \Y & U  & \N & \N &    \Y & \N & \Y &   \N        \\ 
\cesare~\cite{Cesare2014Control}        & 2014 & TDSC       & OM  & S   & P  & F     & \N & \N & C   & \Y & \N & \N & \N &    \Y & \N & \N &   \N        \\
\blex~\cite{blex}                       & 2014 & USENIX Sec & OO  & S   & F  & F     & \N & \Y & \N  & \Y & \N & \N & \N &    \Y & \Y & \N &   \N        \\ 
\cop~\cite{cop, copJournal}             & 2014 & ESEC/FSE   & OO  & S,E & P  & F,B   & \N & \Y & C   & \N & \N & \N & \N &    \Y & \N & \Y &   \N        \\ 
\tedem~\cite{tedem}                     & 2014 & ACSAC      & OM  & S   & B* & B     & \N & \Y & C   & \N & \N & \N & \N &    \Y & \N & \N &   \N        \\ 
\sigmasaed~\cite{sigma}                 & 2015 & DFRWS      & OO  & S   & F  & F     & \N & \N & O   & \N & \N & \N & \N &    \Y & \N & \N &   \D\B      \\ 
\ming~\cite{Ming2015Memoized}           & 2015 & IFIP SEC   & OO  & E   & P  & B     & \N & \Y & I   & \N & \N & \N & \N &    \Y & \Y & \Y &   \B\SR     \\ 
\multimh~\cite{Pewny2015BugSearch}      & 2015 & S\&P       & OM  & S   & B* & B     & \N & \Y & C   & \N & \N & \Y & \Y &    \Y & \N & \Y &   \N        \\ 
\edg~\cite{edg2015}                     & 2015 & SANER      & OO  & I   & F  & I*    & \N & \N & O   & \N & \N & \N & \N &    \Y & \N & \Y &   \B\SR     \\
\discovre~\cite{discovre}               & 2016 & NDSS       & OM  & S   & F  & B     & \N & \N & C   & \Y & \N & \N & \Y &    \Y & \N & \N &   \N        \\ 
\mockingbird~\cite{mockingbird}         & 2016 & SANER      & OM  & S   & F  & F     & \N & \Y & \N  & \N & \N & \N & \Y &    \N & \Y & \N &   \N        \\ 
\esh~\cite{esh}                         & 2016 & PLDI       & OM  & E   & F  & I*    & \N & \Y & \N  & \N & \N & \N & \N &    \Y & \N & \Y &   \N        \\ 
\tpm~\cite{tpm}                         & 2016 & TrustCom   & OO  & S   & P  & F     & \N & \N & \N  & \Y & \N & \N & \N &    \Y & \N & \N &   \N        \\ 
\bindnn~\cite{bindnn}                   & 2016 & SecureComm & OM  & S   & F  & F     & \N & \N & \N  & \N & S  & \N & \Y &    \Y & \N & \N &   \B        \\ 
\genius~\cite{genius}                   & 2016 & CCS        & OM  & S   & F  & B     & \N & \N & C   & \Y & U  & \Y & \Y &    \Y & \N & \N &   \N        \\ 
\bingo~\cite{bingo}                     & 2016 & FSE        & OM  & S   & F  & B*,I* & \N & \Y & \N  & \N & \N & \N & \Y &    \Y & \N & \Y &   \SR       \\ 
\kim~\cite{Kim2016}                     & 2016 & JSCOMPUT   & OO  & S   & P  & F     & \N & \N & G   & \Y & \N & \N & \N &    \Y & \Y & \N &   \N        \\
\kamino~\cite{kam1n0}                   & 2016 & SIGKDD     & OM  & S   & B* & B     & \Y & \N & C   & \Y & \N & \Y & \N &    \Y & \N & \N &   \B        \\ 
\binsequence~\cite{binsequence}         & 2017 & ASIACCS    & OM  & S   & F  & B,I   & \Y & \N & C   & \N & \N & \Y & \N &    \Y & \N & \N &   \B        \\ 
\xmatch~\cite{xmatch}                   & 2017 & ASIACCS    & OM  & S   & F  & I*    & \N & \Y & \N  & \N & \N & \N & \Y &    \Y & \N & \Y &   \N        \\ 
\cacompare~\cite{cacompare}             & 2017 & ICPC       & OM  & S   & F  & F     & \N & \Y & \N  & \N & \N & \Y & \Y &    \Y & \N & \N &   \N        \\ 
\spain~\cite{spain}                     & 2017 & ICSE       & OO  & S,I & P  & F,B   & \Y & \Y & \N  & \N & \N & \N & \N &    \Y & \N & \Y &   \B        \\ 
\binsign~\cite{binsign}                 & 2017 & IFIP SEC   & OM  & S   & F  & F     & \N & \N & \N  & \Y & \N & \Y & \N &    \Y & \N & \N &   \B        \\ 
\gitz~\cite{gitz}                       & 2017 & PLDI       & OM  & E   & F  & I*    & \N & \Y & \N  & \N & \N & \N & \Y &    \Y & \N & \N &   \N        \\ 
\binshape~\cite{binshape}               & 2017 & DIMVA      & OM  & S   & F  & F     & \N & \N & \N  & \Y & \N & \Y & \N &    \Y & \N & \N &   \B        \\ 
\binsim~\cite{binsim}                   & 2017 & USENIX Sec & OO  & S   & T  & I*    & \N & \Y & \N  & \N & \N & \N & \N &    \N & \Y & \Y &   \N        \\ 
\kargen~\cite{Kargen2017}               & 2017 & ASE        & OM  & S   & T  & I*    & \N & \Y & \N  & \Y & \N & \N & \N &    \N & \Y & \N &   \N        \\ 
\imfsim~\cite{imfsim}                   & 2017 & ASE        & OO  & S   & F  & F     & \N & \Y & \N  & \Y & S  & \N & \N &    \N & \Y & \Y &   \N        \\
\Gemini~\cite{gemini}                   & 2017 & CCS        & OM  & S   & F  & F     & \N & \N & C   & \Y & S  & \Y & \Y &    \Y & \N & \N &   \N        \\
\fossil~\cite{fossil}                   & 2018 & TOPS       & OM  & S   & F  & F,B*  & \N & \Y & C   & \Y & \N & \N & \N &    \Y & \N & \N &   \B        \\
\firmup~\cite{firmup}                   & 2018 & ASPLOS     & OM  & E   & F  & I*    & \N & \Y & \N  & \N & \N & \N & \Y &    \Y & \N & \N &   \B        \\
\binarm~\cite{binarm}                   & 2018 & DIMVA      & OM  & S   & F  & F     & \N & \N & C   & \Y & \N & \N & \N &    \Y & \N & \N &   \B        \\ 
\alphadiff~\cite{alphadiff}             & 2018 & ASE        & OO  & S   & P  & F     & \N & \N & \N  & \N & S  & \N & \Y &    \Y & \N & \N &   \N        \\ 
\vulseeker~\cite{vulseeker}             & 2018 & ASE        & OM  & S   & F  & F     & \N & \N & C   & \Y & S  & \N & \Y &    \Y & \N & \Y &   \N        \\ 
\jointlearning~\cite{jointlearning}     & 2019 & BAR        & OM  & S   & B  & B     & \N & \N & \N  & \N & S  & \N & \Y &    \Y & \N & \N &   \B        \\
\innereye~\cite{innereye}               & 2019 & NDSS       & OM  & S   & B* & B     & \N & \N & \N  & \N & S  & \Y & \Y &    \Y & \N & \N &   \B        \\      
\asmtovec~\cite{asm2vec}                & 2019 & S\&P       & OM  & S   & F  & I*    & \N & \N & \N  & \N & S  & \N & \N &    \Y & \N & \N &   \N        \\ 
\safe~\cite{safe2019}                   & 2019 & DIMVA      & OM  & S   & F  & F     & \N & \N & \N  & \N & S  & \N & \Y &    \Y & \N & \N &   \N        \\ 

\hline
                                                         
\end{tabular}
}
\end{table*}

\newcommand*{\RowLeft}[1]{\multicolumn{1}{l|}{#1}}

\begin{table*}[t]
\centering
\scriptsize
\caption{Applications evaluated in the analyzed \bcs research works.
\label{tab:apps}}
{
\setlength\tabcolsep{4pt}
\renewcommand\featuretext[1]{\rotatebox{90}{#1}}
\begin{tabular}{|l|l|} 
\hline
\multicolumn{1}{|c|}{\textbf{Application}} & \multicolumn{1}{c|}{\textbf{Research works}} \\ \hline

Bug Search                      & \tracy~\cite{tracy},
                                  \tedem~\cite{tedem},
                                  \multimh~\cite{Pewny2015BugSearch},
                                  \discovre~\cite{discovre},
                                  \esh~\cite{esh},
                                  \genius~\cite{genius},
                                  \bingo~\cite{bingo},
                                  \binsequence~\cite{binsequence} \\ 
                                & \xmatch~\cite{xmatch}, 
                                  \gitz~\cite{gitz},
                                  \Gemini~\cite{gemini},
                                  \firmup~\cite{firmup},
                                  \binarm~\cite{binarm},
                                  \alphadiff~\cite{alphadiff},
                                  \vulseeker~\cite{vulseeker}\\ \hline

Malware Clustering              & \smit~\cite{smit},
                                                                    \mutantxs~\cite{mutantxs},
                                  \kim~\cite{Kim2016}\\ \hline

Malware Detection               & \kruegel~\cite{Kruegel2005},
                                  \bruschi~\cite{Bruschi2006DSM},
                                  \cesare~\cite{Cesare2014Control}\\ \hline

Malware Lineage                 & \beagle~\cite{beagle},
                                  \iline~\cite{iline},
                                  \ming~\cite{Ming2015Memoized}               \\ \hline

Patch Analysis                  & \flake~\cite{Flake2004Structural},
                                  \dullien~\cite{Dullien2005Graph},
                                  \binhunt~\cite{binhunt},
                                  \mockingbird~\cite{mockingbird},
                                  \binsequence~\cite{binsequence},
                                  \spain~\cite{spain},
                                  \kargen~\cite{Kargen2017} \\ \hline

Patch Generation                & \exediff~\cite{exediff} \\ \hline

Porting Information             & \bmat~\cite{bmat},
                                  \flake~\cite{Flake2004Structural},
                                  \dullien~\cite{Dullien2005Graph} \\ \hline

Software Theft Detection        & \cop~\cite{cop} \\
\hline

\end{tabular}
}
\end{table*}

\section{Applications}
\label{sec:apps}

This section motivates the importance of \bcs by 
describing the applications it enables.
Of the \numpapers papers analyzed, 36 demonstrate an application, 
i.e., present a quantitative evaluation, or case studies, 
of at least one application. 
The other 23 papers present generic \bcs capabilities 
that can be used for multiple applications such as 
binary diffing tools (e.g.,~\cite{bindiff,diaphora,darungrim}), 
binary code search platforms (e.g.,~\cite{bindnn,rendezvous,tpm}), and 
binary clone detection approaches 
(e.g.,~\cite{cacompare,kam1n0,Saebjornsen2009DCC,binclone}).
Table~\ref{tab:apps} summarizes the eight applications identified. 
Most of the 36 papers demonstrate a single application, although a few
(e.g., \flake, \binsequence) 
demonstrate multiple.
One property of an application is whether the application compares
different versions of the same program 
(patch analysis, patch generation, porting information, malware lineage), 
different programs (malware clustering, malware detection, 
software theft detection), 
or can be applied to both cases (bug search).
Next, we detail those applications.

\begin{enumerate}

\item \textbf{Bug search --}
Arguably the most popular application of \bcs
is finding a known bug in a large repository of target 
\chunks~\cite{tedem,Pewny2015BugSearch,discovre,genius,vulseeker,gemini,xmatch,binsequence,bingo,alphadiff,esh,gitz,firmup,binarm,tracy}.
Due to code reuse, the same code may appear in multiple programs,
or even in multiple parts of the same program.
Thus, when a bug is found, it is important to identify similar code that
may have reused the buggy code and contain the same, or a similar, bug.
Bug search approaches take as input a query buggy \chunk and search for 
similar \chunks in a repository. 
A variant of this problem is {\em cross-platform bug search}, 
where the target \chunks in the repository may be compiled for 
different platforms (e.g., x86, ARM, MIPS)~\cite{Pewny2015BugSearch,discovre,genius,gemini,xmatch,bingo,alphadiff,gitz,firmup}. 

\item \textbf{Malware detection --}
\Bcs can be used to detect malware by comparing a given 
executable to a set of previously known malware samples. 
If similarity is high then the input sample is likely a variant of a previously known malware family. 
Many malware detection approaches are purely behavioral, 
comparing system or API call behaviors 
(e.g.,~\cite{forrest1996sense,kirda2006behavior}). 
However, as described in Section~\ref{sec:scope}, 
we focus on approaches that use 
\bcs~\cite{Kruegel2005,Bruschi2006DSM,Cesare2014Control}.

\item \textbf{Malware clustering --}
An evolution of malware detection is 
clustering similar, known malicious, executables into \emph{families}.
Each family cluster should contain executables  
from the same malicious program, 
which can be different versions of the malicious program, 
as well as polymorphic (e.g., packed) variants of a version.
Similar to malware detection, 
we focus on approaches that compare binary code~\cite{smit} 
and exclude purely behavioral approaches 
based on system calls and network traffic
(e.g.,~\cite{Bayer2009Scalable,Perdisci10,firma}).

\item \textbf{Malware lineage --}
Given a set of executables known to belong to the same program, 
lineage approaches build a graph where nodes are program versions and edges 
capture the evolution of the program across versions.
Lineage approaches are most useful with malware because version information is 
typically not available~\cite{beagle,iline,Ming2015Memoized,lineage}.
Since input samples should belong to the same family, 
malware lineage often builds on the results of malware clustering.

\item \textbf{Patch generation and analysis --}
The earliest \bcs application, and one of the most popular, 
is to diff two consecutive, or close, versions of the 
same program to identify what was patched in the newer version.
This is most useful with proprietary programs where the vendor 
does not disclose patch details.
The diffing produces small binary code patches that can be efficiently
shipped to update the program~\cite{exediff}.
It can also be used to automatically identify security patches that 
fix vulnerabilities~\cite{spain}, 
analyze those security patches~\cite{Flake2004Structural,Dullien2005Graph,binhunt,mockingbird,Kargen2017,binsequence}, and  
generate an exploit for the old vulnerable version~\cite{Brumley2008Automatic}.

\item \textbf{Porting information --}
\Bcs can be used for porting information between two 
close versions of the same program. 
Since close versions typically share a large amount of code, 
the analysis done for one version may be largely reusable for a newer version. 
For example, early \bcs approaches ported 
profiling data~\cite{bmat,bmatSPP} and analysis results obtained during 
malware reverse engineering~\cite{Flake2004Structural,Dullien2005Graph}.

\item \textbf{Software theft detection --}
\Bcs can be used for identifying unauthorized reuse of code from 
a plantiff's program such as 
the source code being stolen, its binary code reused, 
a patented algorithm being reimplemented without license,  
or the license being violated (e.g., GPL code in a commercial application).
Early approaches for detecting such infringements used software birthmarks, 
i.e., signatures that capture inherent functionality of the plaintiff 
program~\cite{Myles2005KBS,Choi2007SBB}. 
However, as described in Section~\ref{sec:scope},
we exclude signature-based approaches and focus on approaches using 
\bcs~\cite{cop}.

\end{enumerate}

\section{\BCS Evolution}
\label{sec:evolution}

This section describes the origins of \bcs and its evolution over 
the last \numyears, highlighting some noteworthy approaches.

\paragraph{The origins.}
The origins of \bcs are in the problem of generating a patch (or delta)
that captures the differences between
two consecutive (or close) versions of the same program.
Text diffing tools had existed since the 1970's
(the popular UNIX {\em diff} was released in 1974) and
had been integrated in early source code versioning systems such as
SCCS~\cite{sccs} (1975) and RCS~\cite{rcs} (1985).
The increasing popularity of low bandwidth communication networks
(e.g., wireless)
and the limited resources in some devices,
raised interest in techniques that would increase efficiency by transmitting 
a small patch that captured the differences
between two versions of a binary, i.e., non-text, file,
instead of transmitting the whole file.
In 1991, Reichenberger proposed a diffing technique
for generating patches between arbitrary binary files
without any knowledge about the file structure~\cite{reichenberger1991delta}.
The approach worked at the byte-level,
instead of the line-level granularity of text diffing,
and efficiently identified the
byte sequences that should appear in the patch because they only appeared
in the updated version and thus were not available in the original file
to be patched.
Several tools for generating and applying binary patches soon appeared
such as \rtpatch~\cite{rtpatch},
\bindifforig~\cite{bdiff95}, and
\xdelta~\cite{xdelta}.
Those tools worked at byte-level and could diff any type of file.

The first \bcs approaches are from 1999.
That year, Baker et al. proposed an approach for compressing differences
of executable code and built a prototype diffing tool
called \exediff~\cite{exediff},
which generated patches for DEC Alpha executables.
Their intuition was that many of the changes when diffing two executable
versions of the same program represent secondary changes due to
the compilation process, as opposed to direct changes in the source code.
One example they mentioned is register allocation 
that may change at recompilation.
Another example they mentioned is that code added in the newer version 
would displace parts of the old code, 
and thus the compilation process would have to adjust
pointer values in the displaced code to point to the correct addresses.
Their idea was to reconstruct secondary changes at patch time, 
so that they would not need to be included in the patch, 
reducing the patch size.
\exediff is the earliest approach we have identified
that focused on computing similarity between binary code,
taking advantage of the code structure by disassembling the raw bytes
into instructions. 

Also in 1999, Wang et al. presented \bmat~\cite{bmat}, a tool that aligned two versions of a Windows DLL library executable
to propagate profile information from the older (extensively profiled) version
to a newer version, thus reducing the need for re-profiling.
Their approach is the first to compare functions and basic blocks
(\exediff compared two sequences of instructions). 
It first matched functions in the two executables and then
matched similar blocks within each matched function.
It used a hashing technique to compare blocks.
The hashing removed relocation information to handle pointer changes, 
but was order-sensitive. 

\paragraph{The first decade.}
After the initial works of \exediff and \bmat, 
we only identify 7 \bcs approaches in the next decade (2000-2009). 
However, some of these are highly influential as 
they extend \bcs from purely syntactical to also include semantics; 
they widen the scope from \diffing (OO) to also include 
\clustering (MM) and \search (OM); 
and they apply \bcs to malware.

In 2004, Thomas Dullien (alias Halvar Flake) proposed a 
graph-based \diffing approach that focused on the structural properties 
of the code by heuristically constructing a callgraph isomorphism 
that aligns functions in two versions of the same 
binary program~\cite{Flake2004Structural}.
This is the first approach to handle instruction reordering inside a 
function introduced by some compiler optimizations.
A followup work~\cite{Dullien2005Graph} 
extended the approach to also match basic blocks inside matched functions 
(as done in \bmat)
and introduced the Small Primes Product (SPP) hash 
to identify similar basic blocks despite instruction reordering.
These two works are the basis for the 
popular \bindiff \diffing plugin for the \ida disassembler~\cite{bindiff}. 

In 2005, Kruegel et al. proposed a graph coloring technique to detect
polymorphic variants of a malware. 
This is the first approach that performed semantic similarity 
and MM comparison. 
They categorized instructions with similar functionality into 14 
semantic classes.
Then, they colored the inter-procedural control-flow graph (ICFG) 
using those classes.
The graph coloring is robust against syntactical obfuscations such as 
junk insertion, instruction reordering, and instruction replacement.
In 2008, Gao et al. proposed \binhunt~\cite{binhunt} 
to identify semantic differences between two versions of the same program.
This is the first approach that checks code equivalence. 
It uses symbolic execution and a constraint solver to check if two basic blocks 
provide the same functionality. 

In 2009, Xin et al. presented \smit~\cite{smit}, 
an approach that given a malware sample finds similar malware in 
a repository. \smit is the first OM and \search approach.
It indexes malware callgraphs in a database and uses 
graph edit distance to find malware with similar callgraphs.

\paragraph{The last decade.}
The last decade (2010-2019) has seen a huge increase in the popularity of 
\bcs, with 52 approaches identified. 
The focus on this decade has been on \search approaches,
with an emphasis since 2015 on its cross-architecture version 
(\numcrossarch approaches),
and in recent years on machine learning approaches. In 2013, Wei et al. proposed \rendezvous, 
a \search engine that given the binary code of a query function, 
finds other functions in a repository with similar syntax and 
structural properties.
Reducing the search granularity from whole programs (\smit) to 
smaller \chunks such as functions
enables an array of applications such as clone detection and bug search.

Most \search approaches target the bug search application.
This application was first addressed on source code in 2012 by 
Jang et al.~\cite{redebug}. 
In 2014, David et al. proposed \tracy~\cite{tracy}, 
the first \search approach focused on bug search.
\tracy used the concept of \emph{tracelets}, 
an execution path in a CFG, 
to find functions similar to a vulnerable function.
In 2015, Pewny et al. presented \multimh~\cite{Pewny2015BugSearch}, 
the first cross-architecture \search approach.
\multimh indexed functions by their input-output semantics.
Given a function compiled for one CPU architecture (e.g., x86) 
\multimh can find similar functions compiled for other 
architectures (e.g., MIPS).
This problem quickly gained traction due to the popularity of embedded devices.
In 2016, Lageman et al.~\cite{bindnn} trained a neural network to decide 
if two functions were compiled from the same source code.
The use of deep learning has picked up in the last two years, 
e.g., \alphadiff~(2018), \innereye~(2019), and \asmtovec~(2019).

\section{Approaches}
\label{sec:approach}

In this section, we systematize the approaches of \bcs, 
describing the \textit{Approach Characteristics} columns in 
Table~\ref{tab:approach}.
We recommend the reader to print Table~\ref{tab:approach} in 
a separate page to have it in hand while reading this section. 

\subsection{Comparison Type}
\label{sec:comptype}

\columns{Input Comparison; Approach Comparison}

This section discusses the type of comparison between the approach inputs, 
as well as the finer granularity comparisons that may be used by the approach.

\paragraph{Input comparison.}
All \numapproaches works analyzed compare their inputs to identify similarity. 
That is, no approach identifies identical inputs 
(since a hash suffices for that) or input equivalence 
since it is an undecidable problem~\cite{zakharov01equivalence}  
that can only be solved efficiently for small \chunks.
Thus, we classify \bcs approaches based on their input 
comparison as:
one-to-one (OO, \numoo approaches), one-to-many (OM, \numom approaches), and 
many-to-many (MM, \nummm approaches).
The dominance of $OM$ approaches is due to the high interest in \search  
in the last decade.

It is always possible to build an $OM$ or $MM$ 
approach from an $OO$ approach.
For example, a simple implementation of an $OM$ approach 
is to compare the given query \chunk with each of the $n$ targets
using an $OO$ approach that returns the similarity between both inputs. 
Then, simply rank the $n$ targets by decreasing similarity 
and return the top $k$ entries or the entries above a similarity threshold.
However, most $OM$ approaches avoid this simple implementation 
since it is inefficient.
The two main solutions to improve performance are 
extracting a feature vector from each input and 
storing the target \chunks in a repository with indices. 
Obtaining a feature vector for each input 
allows to perform the feature extraction only once per input.
This offers significant benefits when the feature extraction is expensive,
e.g., in the case of \blex whose feature extraction requires executing 
a \chunk multiple times with different inputs.
Once the feature vectors have been extracted, 
a similarity metric between two feature vectors is used. 
This similarity metric is typically cheap to compute as feature vectors 
are often numerical or Boolean.
A common source of confusion is that some approaches propose similarity 
metrics, while others propose {\em distance} metrics. 
It is important to keep in mind that when the metrics are normalized 
between zero and one, the distance is simply one minus the similarity.
The other solution used by $OM$ approaches is adding indices
on a subset of the features in the feature vector.
The indices can be used to reduce the number of comparisons by 
applying the similarity metric only between the feature vector of 
the input \chunk and the feature vectors of selected targets 
more likely to be similar.

\paragraph{Approach comparison.}
Most approaches use a single type of comparison: 
similarity (42 approaches), equivalence (5), and identical (2).
Note that even if only one type of comparison is used in the approach, 
it may differ from the input comparison. 
For example, \exediff looks for identical instructions in the process of 
diffing two programs.
There are 12 approaches that use multiple comparison types at 
different granularities. 
Of those, six use identical comparison at finer granularities 
to quickly identify the same \chunks
(\bmat, \dullien, \saebjornsen, \binclone)
or to reduce expensive comparisons such as graph isomorphism (\smit, \spain). 
The other six use equivalence comparisons at finer granularities 
to capture semantic similarity.

\subsection{Granularity}
\label{sec:granularity}

\columns{Input Granularity; Approach Granularities}

We separate the input granularity from
the granularities of the \chunks compared in the approach 
(i.e., approach granularities) 
since it is common to use finer granularities (e.g., functions) to 
compare coarser input granularity (e.g., whole programs).

We have identified 8 comparison granularities: 
instruction (I), 
set of related instructions (I*), 
basic block (B), 
set of related basic blocks (B*), 
function (F), 
set of related functions (F*), 
trace (T), and 
whole program (P). 
Instruction, basic block, function, and whole program are 
standard granularities that require no explanation. 
Related instructions (I*) are either consecutive (e.g., n-gram) or 
share a property (e.g., data dependent).
They may belong to different basic blocks, 
and even to different functions.
For example, \tracy groups instructions that collectively 
impact an output variable.
Related basic blocks (B*) 
share structural properties (e.g., graphlets in a CFG) 
or belong to the same execution path.
Basic blocks in a set may belong to the same or multiple functions.
Related functions (F*) implement a program component such as 
a library, a class, or a module. 
Trace granularity compares the execution trace of two binary programs 
on the same input. 

The most common input granularity is 
function (26 approaches) followed by whole program (25) and 
related basic blocks (4). 
Whole program is the preferred input granularity for $OO$ approaches 
(16/\numoo) 
since most \diffing approaches try to establish a one-to-one mapping 
between {\em all} functions in the input programs, 
and also for $MM$ (7/\nummm) approaches that tend to cluster input programs.
On the other hand, function is the preferred granularity 
for \search approaches (21/\numom).  
Another four \search approaches use B* to 
identify code reuse that covers only a subset of a function or   
crosses function boundaries.

The most common approach granularity is function (30 approaches) 
followed by basic block (20).
The majority of approaches (47/\numapproaches) use different input and 
approach granularities, 
i.e., use finer approach granularities to compare coarser input granularity.
Most approaches with the same input and approach granularity perform 
function searches (12/14). 
The 11 approaches that perform equivalence comparisons 
do so at fine granularities due to its low efficiency: 
six have B granularity, 
five I*, and one F. Some approaches accumulate features at a fine granularity that 
are never directly compared and thus do not show in the 
approach granularities column. 
For instance, \Gemini accumulates basic block features
to generate a numerical vector at function granularity. 
Thus, only functions are compared.

\subsection{Syntactic Similarity}
\label{sec:syntax}

\column{Syntactic similarity}

Syntactic approaches capture similarity of the code representation, 
more especifically they compare sequences of instructions. 
Most commonly, the instructions in a sequence are consecutive 
in the virtual address space and belong to the same function.
The instructions in the sequence may first be normalized, 
e.g., considering only the mnemonic, only the opcode, or
normalizing the operands into classes. 
We detail instruction normalization in Section~\ref{sec:normalization} 
and simply refer to instruction sequences in the rest of this subsection.

The instruction sequences may be of fixed or variable length. 
Fixed-size sequences are obtained by sliding a window 
over the instruction stream, 
e.g., over the linearly-ordered instructions in a function.
This process is characterized by the {\em window size},
i.e., the number of instructions in the sequence, 
and the {\em stride}, i.e., the number of instructions to slide the 
start of the window to produce the next sequence. 
When the stride is smaller than the window size, 
consecutive sequences overlap. 
When the stride is one, the resulting sequence is called an n-gram. 
For example, given the sequence of instruction mnemonics
$\{mov, push, add\}$ two 2-grams will be extracted: 
$\{mov,push\}$ and $\{push,add\}$.
There are 7 works that use n-grams: 
\idea, \mbc, \rendezvous, \mutantxs, \expose, \iline, and \kamino.
Fixed-size sequences are also used by \saebjornsen 
with a configurable stride larger than one. 
\rendezvous, in addition to n-grams, also uses n-perms, 
unordered n-grams that capture instruction reordering within the sequence.
An n-perm may capture multiple n-grams, 
e.g., 2-perm $\{mov,push\}$ captures 2-grams $\{mov,push\}$ and $\{push,mov\}$.

The most common methods to compare instruction sequences are  
hashing, embedding, and alignment.
Hashing is used by 6 approaches 
(\bmat, \dullien, \saebjornsen, \smit, \binclone, \spain) 
to obtain a fixed-length value out of a variable-length instruction sequence. 
If the hash values are the same, the sequences are similar. 
Five approaches generate an embedding from n-gram sequences
(\idea, \mbc, \mutantxs, \expose, \kamino).
Three approaches (\exediff, \tracy, \binsequence) align two sequences to 
produce a mapping between them by inserting gaps in either sequence 
to account for inserted, removed, and modified instructions.
These approaches define a similarity score when instructions are aligned, 
and a \emph{gap score} when an instruction aligns with a gap.
Other less common comparison methods are using 
vectors of Boolean features (\iline) and encoding sequences as strings for indexing (\rendezvous).

\subsection{Semantic Similarity}
\label{sec:semantics}

\column{Semantic similarity}

Semantic similarity captures if the code being compared has similar 
effects, as opposed to syntactic similarity that captures 
similarity in the code representation.
The semantics of a \chunk can be described by the changes it produces in the 
process state, i.e., updates to the content of registers and memory. 
We identify 26 approaches computing semantic similarity. 
Most of them capture semantics at basic block granularity 
because basic blocks are straight-line code without control flow. 
Three methods are used to capture semantics: 
instruction classification, 
input-output pairs, and
symbolic formulas. 

\paragraph{Instruction classification.}
The first approach to introduce semantics into \bcs was \kruegel, 
which classified instructions into 14 classes 
(e.g., arithmetic, logic, data transfer) and used a 14-bit value to 
capture the classes of the instructions in a basic block. 
This semantic $color$ bitvector captures the effects of the basic block. 
This approach was later adopted by 
\bruschi, \beagle, \fossil, and \sigmasaed. 
Instruction classification can be used to compute semantic similarity, 
but cannot determine if two \chunks are, or are not, equivalent.

\paragraph{Input-output pairs.}
Intuitively, two \chunks are functionally equivalent if given the same input 
they produce the same output, for all possible inputs. 
Such equivalence is independent of the code representation and 
compares the final state after the code is executed, 
ignoring intermediate program states.
This approach was proposed by Jiang et al. 
on source code~\cite{Jiang2009AMF} and later used by numerous \bcs approaches:
\binhash, \ming, \blex, \multimh, \bingo, \cacompare, \spain, \kargen and \imfsim.
It involves executing both \chunks with the same input and 
comparing their output, repeating the process many times. 
If the output differs for any input, then the two \chunks are not equivalent. 
Unfortunately, to determine that both \schunks are equivalent, the approach 
would require testing all possible inputs, 
which is not realistic for any non-trivial \chunk. 
Thus, in practice, this approach can only determine 
that two \chunks are \emph{likely equivalent}, with a confidence proportional 
to the fraction of inputs that have been tested, or 
that they are not equivalent (with certainty). 
The tested inputs are typically selected randomly, 
although it is possible to use other selection rules, 
e.g., taking values from the program data section (\cacompare).
It is generally a dynamic approach, 
but some approaches (e.g., \multimh) evaluate concrete inputs on 
statically-extracted symbolic formulas.

\paragraph{Symbolic formula.}
A symbolic formula is an assignment statement in which the 
left side is an output variable and the 
right side is a logical expression of input variables and literals 
that captures how to derive the output variable. 
For instance, the instruction {\tt add \%eax,\%ebx} 
can be represented with the symbolic formula
{\tt EBX2 = EAX + EBX1}
where {\tt EBX2} and {\tt EBX1} are symbols representing the values of 
the EBX register before and after executing the instruction.
Eleven approaches use symbolic formulas: 
\binhunt, \ibinhunt, \binhash, \expose, \tracy, \ruttenberg, \tedem,
\cop, \multimh, \esh, and \xmatch.
Of those, eight approaches extract symbolic formulas at basic block granularity, 
\xmatch and \expose extract formulas for the return values of a function, and 
\binsim extracts symbolic formulas from an execution trace that capture how 
the arguments of a system call were derived. 
Three methods are used to compare symbolic formulas:
using a \emph{theorem prover} to check for equivalence,
comparing the \emph{semantic hash} of the formulas to check for equivalence,
and computing the similarity of the graph representation of the formulas.

\noindent
{\it Theorem prover --}
\binhunt introduced the idea of using 
theorem provers such as STP~\cite{stp} or Z3~\cite{z3} 
to check if two symbolic formulas are equivalent, 
i.e., whether the output variable always contains the same value 
after the execution of both formulas, 
assuming that the input variables share the same values.  
The main limitation of this approach 
is that it is computationally expensive because 
it can only perform pairwise equivalence queries, 
the solving time quickly increases with formula sizes, and 
the solver may fail to return an answer for some queries. 
Note that a \chunk may have multiple outputs 
(registers and variables in memory), 
each represented by its own symbolic formula. 
These approaches need to try all pair-wise comparisons 
and check if there exists a permutation of variables 
such that all matched variables contain the same value.

\noindent
{\it Semantic hashes --}
An alternative to using a theorem prover is to check if two symbolic formulas 
have the same hash, after normalizing the formulas 
(e.g., using common register names) and 
simplifying them (e.g., applying constant propagation). 
The intuition is that if the two symbolic formulas have the same hash 
they should be equivalent. 
Three approaches use semantic hashes: \binhash, \binjuice, and \gitz.
Semantic hashes are efficient, but are limited in that 
it is possible for two equivalent formulas to have different hashes 
even after normalization and simplification. 
For example, reordering of symbolic terms in one of the formulas 
(e.g., due to instruction reordering) results in a different hash. 

\noindent
{\it Graph distance --}
\xmatch and \tedem represent the symbolic formula of a basic block as a tree, 
and compute their similarity by applying graph/tree edit distance. 
Computing the graph/tree edit distance is more expensive than 
comparing semantic hashes, but the graph representation has the advantage 
over semantic hashes that it can handle term reordering.

\subsection{Structural Similarity}
\label{sec:structural}

\column{Structural similarity}

Structural similarity computes similarity on graph representations of 
binary code.
It sits between syntactic and semantic similarity since 
a graph may capture multiple syntactic representations of the same code and 
may be annotated with semantic information.
Structural similarity can be computed on different graphs. 
The three most common are the intra-procedural control flow graph (CFG), 
the inter-procedural control flow graph (ICFG), 
and the callgraph (CG). 
All three are directed graphs. 
In the CFG and ICFG, nodes are basic blocks and 
an edge indicates a control flow transition (e.g., branch, jump).
Basic blocks in a CFG belong to a single function; 
each function has its own CFG. 
Basic blocks in the ICFG belong to any program function;   
there is one ICFG per program.
In the CG, nodes are functions and 
edges capture caller-callee relationships.

The intuition behind structural approaches is that CG, ICFG, and CFGs 
are fairly stable representations whose structure varies little 
for similar code. 
Approaches that operate on the CG or ICFG have a 
whole program input granularity, 
while those that operate on CFGs may have function granularity, 
or use function similarity as a step towards whole program similarity.
Structural similarity approaches may use labeled graphs. 
For example, \flake and \dullien use node labels in the CG to capture 
structural information about a function's CFG 
(e.g., number of instructions and edges). 
Other approaches label basic blocks in the CFG/ICFG with a feature vector that 
captures the semantics of the basic clock (e.g., \kruegel, \binjuice)
or its embedding (\genius, \Gemini, see \S\ref{sec:features}).
Edge labels can be used  
to capture the type of control flow transfer (\bruschi) 
or to aggregate the semantic labels of source and destination nodes (\fossil).

Structural similarity is used by 27 approaches.
14 approaches operate only on CFGs;
five on both CFGs and the CG; 
four only on the ICFG; and
two only on the CG.   
There are also three approaches that use non-standard graphs: 
\sigmasaed proposes a \emph{semantic integrated graph} 
that combines information from the CFG, CG, and \emph{register flow graph}, 
while \edg and \libv use the 
\emph{execution dependence graph}~\cite{slavespeculative}.
The remainder of this subsection discusses different approaches 
used to compute graph similarity.

\paragraph{(Sub)Graph isomorphism --}
Most structural similarity approaches check for variations of graph isomorphism.
An isomorphism of two graphs $G$ and $H$ is an edge-preserving bijection $f$
between their node sets such that if any two nodes $u$, $v$ are adjacent in $G$,
then $f(u)$ and $f(v)$ are also adjacent in $H$.
Graph isomorphism requires that the node set cardinality  
is the same in both graphs, which is too strict for \bcs.
Thus, approaches instead check for \emph{subgraph isomorphism}, 
which determines if $G$ contains a subgraph isomorphic to $H$.
Subgraph isomorphism is a known NP-complete problem.
Other approaches check for the \emph{maximum common subgraph isomorphism} (MCS), 
which finds the largest subgraph isomorphic to two graphs and is also 
NP-Complete.
Given the high complexity of both problems, approaches try to reduce the 
number of graph pairs that need to be compared, 
as well as the size of the compared graphs. 
For example, \dullien avoids comparing CFGs with the same hash
(match) and CFGs with very different number of nodes and edges 
(unlikely match).
\ibinhunt reduces the number of nodes to consider 
by assigning taint labels to basic blocks. 
Only nodes with the same taint label are considered in the 
subgraph isomorphism. 
For candidate graph pairs that pass the filtering, 
approximate algorithms are used that 
can be grouped into greedy and backtracking.

\noindent
\textit{Greedy --}
These approaches perform neighborhood exploration.
An initial set of matching nodes is first identified. 
Then, the matching is recursively expanded by checking only the neighbors 
(i.e., parents or children) of already matched nodes. 
\bmat, \flake, \dullien, \lee, \tedem, \multimh, \kim, 
\kamino, \binsequence, and \binarm use this approach.
A limitation of greedy algorithms is that early errors propagate, 
significantly reducing the accuracy.

\noindent
\textit{Backtracking --}
Backtracking algorithms fix a wrong matching by revisiting the solution, 
and if the new matching does not improve the overall matching it is reverted
(\bruschi, \binhunt, \ibinhunt, \ming, \edg, \discovre).
Backtracking is more expensive, 
but can improve accuracy by avoiding local optimal matching. 

\paragraph{Optimization.}
An alternative used by four approaches
(\smit, \binslayer, \cesare, \genius)
is to model graph similarity as an optimization problem. 
Given two CFGs and a \emph{cost function} between two basic blocks, 
they find a bijective mapping between the two CFGs with minimum cost.
Such bipartite matching ignores graph structure,
i.e., does not use edge information. 
To address this, \smit and \binslayer assign lower cost to connected 
basic blocks. 
To perform the matching, 
\smit, \binslayer, and \cesare use the $O(n^3)$ Hungarian algorithm,
while \genius uses a genetic algorithm. 
\paragraph{K-subgraph matching.}
\kruegel proposed to divide a graph into $k$-subgraphs, 
where each subgraph contains only $k$ connected nodes.
Then, generate a fingerprint for each k-subgraph and the 
similarity of two graphs corresponds to the maximum number 
of $k$-subgraphs matched.
Four other approaches later leveraged this approach: 
\beagle, \cesare, \rendezvous, and \fossil. 

\paragraph{Path similarity.}
There are three approaches (\cop, \sigmasaed, \binsequence) 
that convert function similarity into a path similarity comparison. 
First, they extract a set of executions paths from a CFG, 
then define a path similarity metric between execution paths, and 
finally combine the path similarity into a function similarity.

\paragraph{Graph embedding.} 
Another method used by \genius, \vulseeker, and \Gemini, 
detailed in \S\ref{sec:features}, 
is to extract a real-valued feature vector from each graph and then 
compute the similarity of the feature vectors.

\begin{figure*}[t]
\centering
\includegraphics[width=0.85\textwidth]{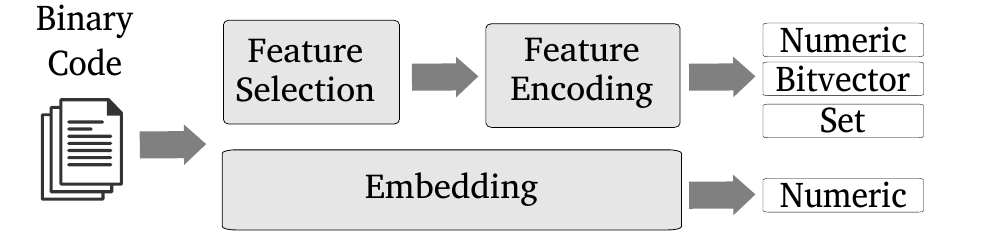}
\caption{Two alternative methods for feature-based similarity.
}
\label{fig:binarycodefeatures}
\vspace{-0.1in}
\end{figure*}

\subsection{Feature-Based Similarity}
\label{sec:features}

\column{Feature-based, Machine learning}

A common method (28 approaches) to compute similarity is to 
represent a \chunk as a vector or a set of features
such that similar \chunks have similar 
\emph{feature vectors} or \emph{feature sets}.
A feature captures a syntactic, semantic, or structural property
of the binary code. 
Features can be Boolean, numeric, or categorical. 
Categorical features have discrete values, 
e.g., the mnemonic of an instruction.
A feature vector typically has all numeric or all Boolean features, 
the latter is called a bitvector. 
Categorical features are typically first encoded into Boolean features using 
one-hot encoding or into real-valued features using an embedding.
Of the 28 approaches, 21 use numeric feature vectors,
six use feature sets, and 
\binclone uses bitvectors.
Once the features have been extracted, 
a similarity metric between feature vectors or feature sets 
is used to compute the similarity. Common similarity metrics are 
the Jaccard index for feature sets, 
dot product for bitvectors, 
and the Euclidean or cosine distance for numeric vectors.

Figure~\ref{fig:binarycodefeatures} shows two alternative methods for 
feature-based similarity. 
The top method comprises of two steps:
\emph{feature selection} and \emph{feature encoding}.
Feature selection is a manual process
where an analyst uses domain knowledge to identify representative features.
The alternative approach showed below is learning to automatically 
generate real-valued feature vectors, 
called \emph{embeddings}, from training data.
Embeddings are used by eight recent approaches 
(\genius, \Gemini, \alphadiff, \vulseeker, \jointlearning, \innereye, \asmtovec, \safe). 
Embeddings are used in natural language processing (NLP) 
to encode categorical features using real-valued numbers, 
which helps deep learning algorithms by reducing the dimensionality 
and increasing the density of feature vectors compared to one-hot encoding.
Embeddings enable automatic feature extraction 
and efficient similarity computation. 
But, features in embeddings do not provide information about what 
has been learnt.

\Bcs embeddings can be classified by the properties they captured 
and their granularity.
The first \bcs approach using embeddings was \genius, 
later followed by \vulseeker and \Gemini.
All the three approaches build a graph embedding for the ACFG of a 
function, i.e., a CFG with nodes annotated with selected basic block features.
While \genius uses clustering and graph edit distance to compute the embedding, 
\vulseeker and \Gemini improve efficiency by training a neural network that avoids 
expensive graph operations.
Later approaches 
(\alphadiff, \jointlearning, \innereye, \asmtovec, \safe)
avoid manually selected features by focusing on instruction, 
or raw byte, co-ocurrencence. 
In NLP, it is common to extract a word embedding that captures word co-occurrence 
(e.g., word2vec) and then build a sentence embedding that builds upon it. 
\jointlearning and \innereye use an analogous approach by considering 
instructions as words and basic blocks as sentences.
\safe uses similar approach to create functions embedding than basic blocks.
Also related is \asmtovec that obtains a function embedding by 
combining path embeddings capturing instruction co-occurrence 
along different execution paths in the function.
Instead of using instruction co-ocurrence, \alphadiff 
computes a function embedding directly from the sequence of raw bytes 
of a function using a convolutional network.

\paragraph{Machine learning.}
We identify three uses of machine learning in \bcs approaches:
(1) to generate an embedding as explained above, 
(2) to cluster similar \chunks using unsupervised learning (\binhash, \mutantxs, \iline, \ruttenberg, \kim, \genius), and
(3) to classify with a probability if the \chunks are being compiled from the same source code 
(\bindnn, \imfsim).
\bindnn is the first use of neural networks for \bcs. 
Instead of generating an embedding,
\bindnn directly uses a neural network classifier to determine if two functions 
are compiled from the same source code. 
Surprisingly, \bindnn is not cited 
by later \bcs approaches including those using neural networks to build embeddings. 

\subsection{Hashing}
\label{sec:hashing}

\column{Locality sensitive hashing}

A hash is a function that maps data of arbitrary size to a fixed-size value. 
Hash values are compact to store, efficient to compute, and 
efficient to compare, which makes them great for indexing. 
Hashes operate at the raw-byte level. 
They are not especifically designed for binary code, 
but rather for arbitrary binary data.   
However, three classes of hashes have been used for \bcs: 
cryptographic hashes, locality-sensitive hashes, and executable file hashes.
Cryptographic hashes capture identical, rather than similar, inputs. 
They are used by some \bcs approaches to quickly identify duplicates at 
fine granularity (e.g., basic block).
Locality-sensitive hashes produce similar hash values for similar inputs, 
as oppossed to cryptographic hashes where a small difference in the input 
generates a completely different hash value.
Executable file hashes take as input an executable file, 
but the hash computation only considers parts of the executable 
such as the import table or selected fields of the executable's header. 
Their goal is to output the same hash value for polymorphic variants 
of the same malware.

\paragraph{Locality sensitive hashing (LSH).}
LSH produces similar hash values for similar inputs, 
efficiently approximating a nearest neighbor search.
LSH algorithms typically apply multiple hash functions on the given input 
and produce the same hash value for similar inputs, 
i.e., they increase collision probability for similar inputs.
LSH is used in \bcs to boost performance. 
For example, it is used by $OM$ approaches for indexing \chunks, 
enabling efficient \search (\kamino, \Gemini, \innereye). 
Of the 11 approaches that use LSH, seven use MinHash~\cite{minhash}
(\binhash, \multimh, \genius, \binsequence, \binshape, \cacompare, \binsign),
two do not specify the algorithm used (\Gemini, \innereye),
\saebjornsen uses the algorithm by Gionis et al.~\cite{gionis1999similarity}, 
and \kamino proposes its own Adaptive Locality Sensitive Hashing (ALSH).

Fuzzy hashing is a popular type of LSH used 
to compute similarity of arbitrary files. 
For example, the VirusTotal file analysis service~\cite{vt} 
reports \emph{ssdeep}~\cite{ssdeep} hashes for submitted files.
Other fuzzy hashes include \emph{\tlsh}~\cite{tlsh},
\emph{\sdhash}~\cite{sdhash}, and 
\emph{\mrsh}~\cite{mrsh-v2}.
None of the \numapproaches approaches use them, 
but we briefly discuss them because 
they are often applied to executable files. 
When applied on executables, fuzzy hashes may capture similarity 
of the binary code, but also similarity of data present in the executables.
This issue has been recently examined by Pagani et al.~\cite{Pagani2018BPR}. 
Their results show that when applied only on the code of the 
\emph{.text} section they work significantly worse than 
when used for whole program similarity. 
In fact, they show that a byte change at the right position, 
extra nop instructions, and instruction swapping
can degrade similarity significantly (in some cases bring it down to zero).
They observe that when compiling the same source code with 
different optimizations, 
the data sections of the executables remain the same, 
which seems to be a key reason fuzzy hashes work better on whole executables.

\paragraph{Executable file hashes.}
This class of hashes are especifically designed for executable files.
They are designed to output the same hash value for 
polymorphic variants of the same malware. 
The hash computation only considers parts of the executable that are less 
likely to change when simply repacking, or resigning, the same executable. 
They are not used by any of the \numapproaches approaches, 
but we briefly describe three popular hashes for completeness. 
\pehash~\cite{pehash} hashes selected fields of a PE executable
that are less susceptible to changes during compilation and packing,
e.g., initial stack size, heap size.
\imphash hashes the import table of an executable. 
Since the functionality of packed variants is the same, their imported 
functions should be the same as well. 
Unfortunately, it gives false positives with unrelated executables packed 
with the same packer, if the packer reconstructs the original import table 
at runtime.
\authentihash is the hash of a PE executable ignoring its 
Windows Authenticode code signing data. 
It enables identifying the same executable signed by different publishers.
It hashes the whole PE executable except three pieces:
the Authenticode data, pointers to the Authenticode data, and 
the file checksum.

\subsection{Supported Architectures}
\label{sec:crossarch}

\column{Cross-architecture}

A cross-architecture approach can compare \chunks for 
different CPU architectures, e.g., x86, ARM, and MIPS.
This differs from architecture-independent approaches (e.g., \flake) 
that support different architectures, 
but cannot cross-compare among them, 
i.e., they can compare two x86 inputs and two MIPS inputs, 
but cannot compare an x86 input with a MIPS input.
There are \numcrossarch cross-architecture approaches, all proposed since 2015.
A common application is given a buggy \chunk,
to search for similar \chunks, compiled for other architectures, 
which may also contain the bug. 
For example, to search for programs in firmware images 
where a version of OpenSSL vulnerable to Heartbleed 
has been statically compiled.

The code syntax for different architectures may significantly differ as 
they may use separate instruction sets with different instruction mnemonics, 
sets of registers, and default calling conventions. 
Thus, cross-architecture approaches compute semantic similarity.
Cross-architecture approaches employ one of two techniques.
Seven approaches lift the 
binary code to an architecture-independent intermediate representation (IR):
\multimh, \mockingbird, \bingo, \xmatch, \cacompare, \gitz, \firmup.
Then, identical analysis can be peformeed on the IR, 
regardless of the origin architecture.  
The advantage is that the analysis only depends on the IR and the IR design 
can be outsourced to a separate group. 
Section~\ref{sec:implementation} details the specific architectures 
supported by each approach and the IRs they use.  
An alternative approach used by 9 approaches is to 
use feature-based similarity (discussed in \S\ref{sec:features}). 
These approaches use a separate module for each architecture to obtain 
a feature vector that captures the semantics of the binary code 
(\discovre, \bindnn, \genius, \Gemini, \alphadiff, \vulseeker, \jointlearning,
\innereye, \safe).

\subsection{Type of Analysis}
\label{sec:typeofanalysis}

\column{Static analysis; Dynamic analysis; Dataflow analysis}

\Bcs approaches can use static analysis, dynamic analysis, or both. 
Static analysis examines the disassembled binary code, without executing it. 
Instead, dynamic analysis examines code executions by running the 
code on selected inputs. 
Dynamic analysis can be performed online, 
as the code executes in a controlled environment, 
or offline on traces of the execution. 
Fifty one approaches use only static analysis, 
four use only dynamic analysis, and six combine both.
The dominance of static analysis for \bcs is due to most applications 
requiring all the input code to be compared. 
This is easier with static analysis as it provides complete code coverage. 
Dynamic analysis examines one execution at a time and 
can only determine similarity of the code run in that execution. 
To increase the coverage of dynamic analysis, 
three approaches (\ibinhunt, \blex, \imfsim) run the code on multiple inputs 
covering different execution paths and combine the results.
However, it is infeasible to run any non-trivial \chunk 
on inputs that cover all possible execution paths.

One advantage of dynamic analysis is simplicity, 
as memory addresses, operand values, and control-flow targets 
are known at runtime, which sidesteps static analysis challenges such as 
memory aliasing and indirect jumps. 
Another advantage is that it can handle some obfuscations and does not 
require disassembly of the code, 
also difficult with obfuscated code~\cite{oec2003}. 
Section~\ref{sec:evaluation} details which approaches have evaluated its 
robustness on obfuscated code. 
Overall, dynamic analysis has been used in \bcs for 
malware unpacking~(\smit, \beagle, \mutantxs, \cesare, \binsim), 
for operating on trace granularity (\binsim, \kargen), and for
collecting runtime values for semantic similarity (\blex, \binsim, \imfsim).

Dataflow analysis is a common type of analysis that
examines how values propagate through the code. 
It comprises of {\em data sources} to be tracked 
(e.g., registers or memory locations holding specific variables), 
{\em propagation rules} defining how values are propagated by 
different instructions or IR statements, and {\em sinks}, 
i.e., program points where to check the values reaching them.
Of the 19 approaches that use dataflow analysis, 
16 use symbolic execution to extract a set of symbolic formulas 
to compute semantic similarity (\S\ref{sec:semantics}).
\spain uses taint analysis to summarize the patterns of 
vulnerabilities and their security patches. 
And, \ibinhunt uses both taint analysis and symbolic execution.
It first uses taint analysis as a filter to find \chunks that process the 
same user input, restricting the expensive subgraph isomorphism computation 
to those with the same taint label. 
And, it computes basic block similarity using symbolic formulas.
\imfsim uses backward taint analysis to infer pointer
arguments of a function from dereference errors.

\subsection{Normalization}
\label{sec:normalization}

\column{Normalization}

Syntactic similarity approaches often normalize instructions, 
so that two instructions that are normalized to the same form 
are considered similar despite some syntactic differences, 
e.g., different registers being used. 
Overall, there are 33 approaches that use instruction normalization.
They apply the following three types of instruction normalization: 

\begin{itemize}

\item \textbf{Operand removal --}
A normalization used by nine approaches is to abstract an 
instruction only by its mnemonic or opcode, ignoring all operands.
For example, $add \, \%eax, \%ebx$ and $add \, [\%ecx], \%edx$ would be both 
represented by $add$ and considered similar, despite both using 
different operands.  

\item \textbf{Operand normalization --}
A normalization used by 17 approaches is to replace 
instruction operands with symbols that capture the operand type
such as \emph{REG} for register, \emph{MEM} for memory, and 
\emph{IMM} for immediate values. For example, $add \, \%eax, \%ebx$ and $add \, \%ecx, \%edx$ would be both 
represented as $add \, REG, REG$, 
matching the instructions despite different register allocations 
used by the compiler. Operand normalization abstracts less than operand removal. 
For example, $add \, [\%ecx], \%edx$ would be represented
as $add \, MEM, REG$ and thus be considered different from the above.
Some approaches also use different symbols for 
general purpose registers and segment registers, and 
for operands of different sizes (e.g., RegGen8 and RegGen32 in \binclone). 

\item \textbf{Mnemonic normalization --}
A normalization used by 3 approches (\bmat, \expose and \iline) is to 
represent multiple mnemonics by the same symbol. 
For example, both \bmat and \iline represent all conditional jumps 
(e.g., {\tt je}, {\tt jne}) with the same symbol to account for the compiler 
modifying jump conditions.
\end{itemize}

Another type of normalization is to ignore code that should not 
affect the semantics. 
For example, some instruction sets contain no-op instructions 
that do not change the process state. 
And, compilers often use no-op equivalent instructions for padding 
such as instructions that move a register to itself, 
e.g., $mov \, \%eax, \%eax$.
No-op equivalent instructions do not matter for semantic similarity and 
structural similarity approaches, 
but since they change the syntax they may affect syntactic 
similarity approaches. 
Three approaches remove no-op instructions~(\iline, \ming, \edg). 
A few approaches also remove unreachable dead code~(\bruschi, \firmup), 
which may be introduced by obfuscations;
function epilogue and prologue~(\expose) instructions,
which may make small unrelated functions to look similar; and 
functions added by the compiler to load the program, 
not present in the source code (\bingo, \sigmasaed).

\begin{table*}
\centering
\scriptsize
\caption{Comparison of the implementations of \bcs approaches. 
Symbol -- means information is unknown. 
\label{tab:implementation}}
{
\setlength\tabcolsep{4pt}
\renewcommand\featuretext[1]{\rotatebox{90}{#1}}
\begin{tabular}{|r|l|l|l|l|l||c|ll|lll|lll||ll|}  \hline
& \multicolumn{2}{c|}{} & & & & \multicolumn{9}{c||}{\textbf{Targeted Arch. \& OS}} & \multicolumn{2}{c|}{\textbf{Release}} \\
\cline{7-17}

& \multicolumn{2}{c|}{\textbf{Platform}} & & & & & \multicolumn{2}{c|}{\textbf{CISC}} & \multicolumn{3}{c|}{\textbf{RISC}} &  \multicolumn{3}{c||}{\textbf{OS}} & \multicolumn{2}{c|}{} \\
\cline{2-3} 

\textbf{\system} &

\textbf{Static} &
\textbf{Dynamic} &

\textbf{Prog. Lang.} &
\featuretext{\textbf{Distributed}} &
\featuretext{\textbf{Uses IR}} &
\featuretext{\textbf{Firmware}} &

\featuretext{\textbf{x86}} &
\featuretext{\textbf{x86-64}} &
\featuretext{\textbf{ARM}} &
\featuretext{\textbf{MIPS}} &
\featuretext{\textbf{Other}} &

\featuretext{\textbf{Windows}} &
\featuretext{\textbf{Linux}} &
\featuretext{\textbf{MacOS}} &

\featuretext{\textbf{Open source}} &
\featuretext{\textbf{Free binary}} \\
\hline
\exediff~\cite{exediff}                  & --           & \N       & --             & \N & \N & \N &   \N & \N   & \N & \N & \Y &   \N & \Y & \N &   \N & \N \\
\bmat~\cite{bmat}                        & Vulcan       & \N       & --             & \N & \N & \N &   \Y & \N   & \N & \N & \N &   \Y & \N & \N &   \N & \N \\
\flake~\cite{Flake2004Structural}        & IDA          & \N       & --             & \N & \N & \N &   \Y & \N   & \N & \N & \N &   \Y & \Y & \N &   \N & \N \\
\dullien~\cite{Dullien2005Graph}         & IDA          & \N       & --             & \N & \N & \N &   \Y & \N   & \N & \N & \N &   \Y & \N & \N &   \N & \Y \\
\kruegel~\cite{Kruegel2005}              & --           & \N       & --             & \N & \N & \N &   \Y & \N   & \N & \N & \N &   \Y & \Y & \N &   \N & \N \\
\bruschi~\cite{Bruschi2006DSM}           & Boomerang    & \N       & --             & \N & \N & \N &   \Y & \N   & \N & \N & \N &   \N & \Y & \N &   \N & \N \\
\binhunt~\cite{binhunt}                  & IDA          & \N       & --             & \N & \Y & \N &   \Y & \N   & \N & \N & \N &   \Y & \Y & \N &   \N & \N \\
\saebjornsen~\cite{Saebjornsen2009DCC}   & IDA          & \N       & C++            & \N & \Y & \N &   \Y & \N   & \N & \N & \N &   \Y & \Y & \N &   \N & \N \\
\smit~\cite{smit}                        & IDA          & \N       & C++            & \N & \N & \N &   \Y & \N   & \N & \N & \N &   \Y & \N & \N &   \N & \N \\
\idea~\cite{idea}                        & NewBasic     & \N       & --             & \N & \N & \N &   \Y & \N   & \N & \N & \N &   \Y & \N & \N &   \N & \N \\
\mbc~\cite{mbc}                          & --           & \N       & --             & \N & \N & \N &   \Y & \N   & \N & \N & \N &   \Y & \N & \N &   \N & \N \\
\ibinhunt~\cite{ibinhunt}                & IDA          & Temu     & --             & \N & \Y & \N &   \Y & \N   & \N & \N & \N &   \N & \Y & \N &   \N & \N \\
\beagle~\cite{beagle}                    & --           & Anubis   & --             & \N & \N & \N &   \Y & \N   & \N & \N & \N &   \Y & \N & \N &   \N & \N \\
\binhash~\cite{binhash}                  & ROSE         & \N       & --             & \N & \N & \N &   \Y & \N   & \N & \N & \N &   \Y & \N & \N &   \N & \N \\
\binjuice~\cite{binjuice}                & IDA          & \N       & Python, Prolog & \N & \N & \N &   \Y & \N   & \N & \N & \N &   \Y & \Y & \N &   \N & \N \\
\binslayer~\cite{binslayer}              & DynInst      &          & C++            & \N & \N & \N &   \Y & \N   & \N & \N & \N &   \N & \Y & \N &   \Y & \N \\
\rendezvous~\cite{rendezvous}            & DynInst      & \N       & C++            & \N & \N & \N &   \Y & \N   & \N & \N & \N &   \N & \Y & \N &   \N & \N \\
\mutantxs~\cite{mutantxs}                & IDA          & \N       & --             & \N & \N & \N &   \Y & \N   & \N & \N & \N &   \N & \Y & \N &   \N & \N \\
\expose~\cite{expose}                    & IDA          & \N       & --             & \N & \N & \N &   \Y & \N   & \N & \N & \N &   \N & \Y & \N &   \N & \N \\
\iline~\cite{iline}                      & IDA          & PIN      & C, Python      & \N & \N & \N &   \Y & \N   & \N & \N & \N &   \Y & \Y & \N &   \N & \N \\
\lee~\cite{Lee2013}                      & IDA          & \N       & Python         & \N & \N & \N &   \Y & \N   & \N & \N & \N &   \Y & \N & \N &   \N & \N \\
\tracy~\cite{tracy}                      & IDA          & \N       & Python         & \N & \N & \N &   \Y & \Y   & \N & \N & \N &   \N & \Y & \N &   \Y & \N \\
\binclone~\cite{binclone}                & IDA          & \N       & C++            & \N & \N & \N &   \Y & \N   & \N & \N & \N &   \Y & \N & \N &   \N & \N \\
\ruttenberg~\cite{Ruttenberg2014Shared}  & IDA          & \N       & Python         & \N & \N & \N &   \Y & \N   & \N & \N & \N &   \Y & \N & \N &   \N & \N \\
\cesare~\cite{Cesare2014Control}         & Malwise      & \N       & C++            & \N & \N & \N &   \Y & \N   & \N & \N & \N &   \Y & \N & \N &   \N & \N \\ \blex~\cite{blex}                        & IDA          & PIN      & C++            & \N & \N & \N &   \N & \Y   & \N & \N & \N &   \N & \Y & \N &   \N & \N \\
\cop~\cite{cop}                          & IDA, BAP     & \N       & C++            & \N & \Y & \N &   \Y & \N   & \N & \N & \N &   \N & \Y & \N &   \N & \N \\
\tedem~\cite{tedem}                      & IDA          & \N       & C++            & \N & \Y & \N &   \Y & \N   & \N & \N & \N &   \Y & \Y & \Y &   \N & \N \\
\sigmasaed~\cite{sigma}                  & --           & \N       & --             & \N & \N & \N &   \Y & \N   & \N & \N & \N &   \Y & \N & \N &   \N & \N \\
\ming~\cite{Ming2015Memoized}            & \N           & Temu     & Ocaml          & \N & \Y & \N &   \Y & \N   & \N & \N & \N &   \Y & \N & \N &   \N & \N \\
\multimh~\cite{Pewny2015BugSearch}       & IDA          & \N       & C++            & \N & \Y & \Y &   \Y & \N   & \Y & \Y & \N &   \Y & \Y & \Y &   \N & \N \\
\edg~\cite{edg2015}                      & IDA          & \N       & C++            & \N & \N & \N &   \Y & \N   & \N & \N & \N &   \Y & \N & \N &   \Y & \N \\ \discovre~\cite{discovre}                & IDA          & \N       & --             & \N & \N & \Y &   \Y & \N   & \Y & \Y & \N &   \Y & \Y & \N &   \N & \N \\
\mockingbird~\cite{mockingbird}          & IDA          & Valgrind & Python         & \N & \Y & \N &   \Y & \N   & \Y & \Y & \N &   \N & \Y & \N &   \N & \N \\
\esh~\cite{esh}                          & IDA, BAP     & \N       & C\#, Python    & \N & \Y & \N &   \N & \Y   & \N & \N & \N &   \N & \Y & \N &   \Y & \N \\
\tpm~\cite{tpm}                          & IDA          & \N       & Python         & \N & \N & \N &   \Y & \N   & \N & \N & \N &   \Y & \N & \N &   \N & \N \\
\bindnn~\cite{bindnn}                    & IDA          & \N       & --             & \N & \N & \N &   \Y & \Y   & \Y & \N & \N &   \N & \Y & \N &   \N & \N \\
\genius~\cite{genius}                    & IDA          & \N       & Python         & \N & \N & \Y &   \Y & \N   & \Y & \Y & \N &   \N & \Y & \N &   \Y & \N \\
\bingo~\cite{bingo}                      & IDA          & \N       & Python         & \N & \Y & \N &   \Y & \Y   & \Y & \N & \N &   \Y & \Y & \Y &   \N & \N \\
\kim~\cite{Kim2016}                      & IDA          & PIN      & --             & \N & \N & \N &   \Y & \N   & \N & \N & \N &   \Y & \N & \N &   \N & \N \\
\kamino~\cite{kam1n0}                    & IDA          & \N       & --             & \Y & \N & \N &   \Y & \Y   & \N & \N & \N &   \Y & \N & \N &   \Y & \N \\  
\binsequence~\cite{binsequence}          & IDA          & \N       & --             & \N & \N & \N &   \N & \Y   & \N & \N & \N &   \Y & \N & \N &   \N & \N \\ \xmatch~\cite{xmatch}                    & IDA, McSema  & \N       & --             & \N & \Y & \Y &   \Y & \N   & \N & \Y & \N &   \Y & \Y & \Y &   \N & \N \\
\cacompare~\cite{cacompare}              & IDA          & \N       & Python         & \N & \Y & \N &   \Y & \N   & \Y & \Y & \N &   \N & \Y & \N &   \N & \N \\
\spain~\cite{spain}                      & IDA          & \N       & --             & \N & \N & \N &   \Y & \N   & \N & \N & \N &   \Y & \Y & \N &   \N & \N \\
\binsign~\cite{binsign}                  & IDA          & \N       & --             & \Y & \N & \N &   \Y & \N   & \N & \N & \N &   \Y & \N & \N &   \N & \N \\
\gitz~\cite{gitz}                        & BAP, McSema  & \N       & --             & \N & \Y & \N &   \N & \Y   & \Y & \N & \N &   \N & \Y & \N &   \N & \N \\
\binshape~\cite{binshape}                & IDA          & \N       & Python         & \N & \N & \N &   \N & \Y   & \N & \N & \N &   \Y & \Y & \N &   \N & \N \\
\binsim~\cite{binsim}                    & \N           & Temu     & --             & \N & \Y & \N &   \Y & \N   & \N & \N & \N &   \Y & \N & \N &   \N & \N \\
\kargen~\cite{Kargen2017}                & \N           & PIN      & C++, Python    & \N & \N & \N &   \Y & \Y   & \N & \N & \N &   \N & \Y & \N &   \N & \N \\
\imfsim~\cite{imfsim}                    & \N           & PIN      & C++, Python    & \N & \N & \N &   \N & \Y   & \N & \N & \N &   \N & \Y & \N &   \N & \N \\
\Gemini~\cite{gemini}                    & IDA          & \N       & Python         & \N & \N & \Y &   \Y & \N   & \Y & \Y & \N &   \Y & \Y & \N &   \Y & \N \\
\fossil~\cite{fossil}                    & IDA, Dyninst & \N       & Python         & \N & \N & \N &   \Y & \N   & \N & \N & \N &   \Y & \Y & \N &   \N & \N \\
\firmup~\cite{firmup}                    & IDA, Angr    & \N       & --             & \N & \Y & \Y &   \Y & \N   & \Y & \Y & \Y &   \N & \Y & \N &   \N & \N \\ \binarm~\cite{binarm}                    & IDA          & \N       & C++, Ptyhon    & \N & \N & \Y &   \N & \N   & \Y & \N & \N &   \N & \Y & \N &   \N & \N \\
\alphadiff~\cite{alphadiff}              & IDA          & \N       & --             & \N & \N & \Y &   \Y & \Y   & \Y & \Y & \N &   \N & \Y & \N &   \N & \N \\
\vulseeker~\cite{vulseeker}              & IDA          & \N       & Python         & \N & \N & \Y &   \Y & \Y   & \Y & \Y & \N &   \N & \Y & \N &   \Y & \N \\
\jointlearning~\cite{jointlearning}      & --           & \N       & --             & \N & \N & \N &   \N & \Y   & \Y & \N & \N &   \N & \Y & \N &   \Y & \N \\
\innereye~\cite{innereye}                & BAP          & \N       & Python         & \N & \N & \N &   \Y & \N   & \Y & \N & \N &   \N & \Y & \N &   \Y & \N \\
\asmtovec~\cite{asm2vec}                 & IDA          & \N       & --             & \N & \N & \N &   \Y & \Y   & \N & \N & \N &   \N & \Y & \N &   \Y & \N \\
\safe~\cite{safe2019}                    & IDA, ANGR    & \N       & Python         & \N & \N & \N &   \N & \Y   & \Y & \N & \N &   \N & \Y & \N &   \Y & \N \\

\hline                                                                        
\end{tabular}
}
\end{table*}

\section{Implementations}
\label{sec:implementation}

This section systematizes the implementation of the \numapproaches approaches.
For each approach, Table~\ref{tab:implementation} shows 
the static and dynamic platforms it builds on, 
the programming language used to implement the approach, 
whether the implementation supports distributing the analysis, 
the supported target program architectures and operating systems, and 
how the approach is released.
A dash (-) indicates that we could not find the information (i.e., unknown), 
while a cross (\N) means unsupported/unused.

Building a \bcs approach from scratch requires significant effort.
Thus, all approaches build on top of previously
available binary analysis platforms or tools, which provide functionality
such as disassembly and control flow graphs for static analysis, or
instruction-level monitoring for dynamic analysis.
However, the implementation of an approach may only use
a subset of the functionality offered by the underlying platform.
The most popular static platform is \ida (42 approaches),
followed by BAP~(4), DynInst~(3), and McSema~(2).
\ida main functionalities are disassembly and building control flow graphs. 
Its popularity comes from supporting a large number of architectures.
Some binary analysis platforms already support using \ida as their
disassembler, so it is not uncommon to combine \ida and another platform 
(6 approaches).
Among dynamic approaches, 
\pin is the most popular with 5 approaches,
followed by \temu and \anubis with 3 approaches each.
Previous work has analyzed binary analysis platforms used by binary code 
type inferece approaches~\cite{types}. 
Since most platforms overlap, 
we refer the reader to that work for platform details, 
but provide an extended version of their table in the Appendix 
(Table~\ref{tab:platforms}) with six extra platforms.

A few approaches build on top of previous \bcs approaches. 
One case is that both approaches have overlapping authors. 
For example, \ruttenberg is based on \binjuice, 
and \ming extends \ibinhunt, which already shared components with \binhunt.
The other case is using a previously released approach. 
For example, \binslayer and \spain use \bindiff as a first stage filter to 
find matched functions.

The most popular programming language is Python (20 approaches), 
followed by C++ (14). 
One reason for this is that the majority of approaches use \ida for disassembly 
and \ida supports both C++ and Python plugins.
Moreover, two approaches (\kamino, \binsign) have used distributed
platforms, such as Hadoop, to distribute their analysis.

Binary code analysis can operate directly on a particular instruction set
(e.g., x86, ARM) or convert the instruction set into an 
intermediate representation (IR).
Using an IR has two main advantages.
First, it is easier to reuse the analysis built on top of an IR. 
For example, supporting a new architecture only requires adding a new front-end 
to translate into the IR, but the analysis code can be reused. 
Second, complex instruction sets (e.g., x86/x86-64) can be converted 
into a smaller set of IR statements and expressions, 
which make explicit any side-effects such as implicit operands or 
conditional flags.
There are 15 approaches that use an IR and most use the IR 
provided by the underlying platform. 
Out of 15 approaches, 5 use VINE provided by \bitblaze 
(\binhunt, \ibinhunt, \cop, \ming, \binsim), 
another 5 use VEX provided with \valgrind 
(\multimh, \mockingbird, \cacompare, \gitz, \firmup), and 
two use LLVM-IR (\esh, \xmatch). 
The remaining three approaches use
SAGE III (\saebjornsen), METASM (\tedem) and REIL (\bingo). 
It is surprising that only 6 of the \numcrossarch cross-architecture approaches use an IR
(\multimh, \xmatch, \mockingbird, \cacompare, \gitz, \firmup). 
The remaining approaches provide separate analysis modules for each architecture
that typically output a feature vector with a common format.

The most supported architectures for the target programs to be analyzed 
are x86/x86-64 (59 approaches), 
followed by ARM (16), and MIPS (10). 
The only two approaches that do not support x86/x86-64 are \exediff 
that targets Digital Alpha, and \binarm that targets ARM.
There are \numcrossarch cross-architecture approaches and 9 
approaches that support firmware.
The first cross-architecture approach was \multimh, 
which added support for ARM in 2015. 
Since then, ARM support has become very prevalent due to the popularity 
of mobile and IoT devices.
It is worth noting that even if there are 42 approaches that use \ida,
which supports more than 60 processor families, 
most approaches built on top of \ida only analyze x86/x86-64 programs.
The most supported OS is Linux (41 approaches) 
followed by Windows (35). 
Only 4 approaches support MacOS. 
Early approaches that focused on x86/x86-64 often used \ida to obtain support 
for both PE/Windows and ELF/Linux executables.
Most recently, all approaches that leverage ARM support Linux, 
which is used by Android and also by many IoT devices.

Of the \numapproaches approaches, only 12 are open source
(\binslayer, \tracy, \edg, \esh, \genius, \kamino, \Gemini, \asmtovec,
\vulseeker, \jointlearning, \innereye, \safe).
\dullien was implemented in the \bindiff commercial tool, 
which is now available as a free binary.
The remaining approaches have not been released in any form,
although the platforms they build on may be open source.
Moreover, 4 of the 12 open-source approaches~(\esh, \genius, \jointlearning, \innereye)
have partially released their source code and machine learning models.

\section{Evaluations}
\label{sec:evaluation}

This section systematizes the evaluation of the \numapproaches \bcs approaches.
For each approach, Table~\ref{tab:evaluation} summarizes the 
datasets used (\S\ref{sec:datasets}) and 
the evaluation methodology~(\S\ref{sec:evalmethods}). 

\begin{table*}
\centering
\scriptsize
\caption{Comparison of the evaluations of \bcs approaches.
On the left it summarizes the datasets used and on the right the 
evaluation methodology.
A dash (--) means we could not find the information.
For Boolean columns, \N~indicates no support.
\label{tab:evaluation}
}
{
\setlength\tabcolsep{4pt}
\renewcommand\featuretext[1]{\rotatebox{90}{#1}}
\begin{tabular}{|r||rrr|rr||cccccccc|crc|}  \hline
 & \multicolumn{5}{c||}{\textbf{Datasets}} & \multicolumn{11}{c|}{\textbf{Methodology}} \\
\cline{2-17}

\textbf{\system} & 
\featuretext{\textbf{Executables}} &
\featuretext{\textbf{Benign}} &
\featuretext{\textbf{Malware}} & 

\featuretext{\textbf{Functions}} & 
\featuretext{\textbf{Firmwares}} &

\featuretext{\textbf{GCC}} &
\featuretext{\textbf{ICC}} &
\featuretext{\textbf{MSVC}} &
\featuretext{\textbf{Clang}} &
\featuretext{\textbf{Cross-optimization}} &
\featuretext{\textbf{Cross-compiler}} &
\featuretext{\textbf{Cross-architecture}} &
\featuretext{\textbf{Obfuscation}} & 

\featuretext{\textbf{Accuracy}} &

\featuretext{\textbf{Comparison}} &
\featuretext{\textbf{Performance}} \\ \hline 

\exediff~\cite{exediff}                 &      38 &      38 &       0 &       0 &      0 & \Y & \N & \N & \N  & \N & \N & \N &     \N &    \N &    1 &  \N  \\ 
\bmat~\cite{bmat}                       &      32 &      32 &       0 &       0 &      0 & \N & \N & \Y & \N  & \N & \N & \N &     \N &    \Y &    0 &  \Y  \\ 
\flake~\cite{Flake2004Structural}       &       8 &       8 &       0 &       0 &      0 & \N & \N & \Y & \N  & \N & \N & \N &     \N &    \N &    1 &  \Y  \\ 
\dullien~\cite{Dullien2005Graph}        &       6 &       4 &       2 &       0 &      0 & \N & \N & \Y & \N  & \N & \N & \N &     \N &    \N &    0 &  \Y  \\ 
\kruegel~\cite{Kruegel2005}             &     503 &      61 &     442 &       0 &      0 & \Y & \N & \Y & \N  & \N & \N & \N &     \Y &    \Y &    0 &  \N  \\ 
\bruschi~\cite{Bruschi2006DSM}          &     587 &     572 &     115 &       0 &      0 & \N & \N & \N & \N  & \N & \N & \N &     \N &    \Y &    0 &  \Y  \\ 
\binhunt~\cite{binhunt}                 &       6 &       6 &       0 &       0 &      0 & \Y & \N & \Y & \N  & \N & \N & \N &     \N &    \Y &    0 &  \Y  \\ 
\saebjornsen~\cite{Saebjornsen2009DCC}  &      -- &   1,723 &       0 &       0 &      0 & \Y & \N & \Y & \N  & \Y & \N & \N &     \N &    \Y &    0 &  \Y  \\ 
\smit~\cite{smit}                       & 102,391 &       0 & 102,391 &       0 &      0 & \N & \N & \Y & \N  & \N & \N & \N &     \Y &    \Y &    0 &  \Y  \\ 
\idea~\cite{idea}                       &  26,189 &  13,000 &  13,189 &       0 &      0 & \N & \N & \Y & \N  & \N & \N & \N &     \N &    \N &    0 &  \N  \\ 
\ibinhunt~\cite{ibinhunt}               &       9 &       9 &       0 &       0 &      0 & \Y & \N & \N & \N  & \N & \N & \N &     \N &    \N &    1 &  \Y  \\ 
\mbc~\cite{mbc}                         &      27 &       0 &      27 &       0 &      0 & \N & \N & \Y & \N  & \N & \N & \N &     \N &    \N &    0 &  \N  \\ 
\beagle~\cite{beagle}                   &     381 &       0 &     381 &       0 &      0 & \N & \N & \Y & \N  & \N & \N & \N &     \N &    \Y &    1 &  \N  \\ 
\binhash~\cite{binhash}                 &      16 &       0 &      16 &      -- &      0 & \N & \N & \Y & \N  & \N & \N & \N &     \N &    \N &    0 &  \N  \\ 
\binjuice~\cite{binjuice}               &      70 &      20 &      50 &       0 &      0 & \Y & \N & \Y & \N  & \N & \N & \N &     \N &    \N &    0 &  \N  \\ 
\binslayer~\cite{binslayer}             &      44 &      44 &       0 &       0 &      0 & -- & -- & -- & --  & \N & \N & \N &     \N &    \Y &    1 &  \Y  \\ 
\rendezvous~\cite{rendezvous}           &      98 &      98 &       0 & 0.004 M &      0 & \Y & \N & \N & \Y  & \Y & \Y & \N &     \N &    \Y &    0 &  \Y  \\ 
\mutantxs~\cite{mutantxs}               & 137,055 &       0 & 137,055 &       0 &      0 & \N & \N & \Y & \Y  & \N & \N & \N &     \Y &    \Y &    0 &  \Y  \\ 
\expose~\cite{expose}                   &   3,075 &   3,075 &       0 &       0 &      0 & \Y & \N & \N & \N  & \N & \N & \N &     \N &    \Y &    0 &  \Y  \\ 
\iline~\cite{iline}                     &   1,891 &   1,777 &     114 &       0 &      0 & \Y & \N & \Y & \N  & \N & \N & \N &     \N &    \Y &    0 &  \Y  \\ 
\lee~\cite{Lee2013}                     &      20 &      16 &       4 &       0 &      0 & \N & \N & \Y & \N  & \N & \N & \N &     \N &    \Y &    1 &  \N  \\ 
\tracy~\cite{tracy}                     &      -- &      -- &       0 &   1.0 M &      0 & \Y & \N & \N & \N  & \N & \N & \N &     \N &    \Y &    0 &  \Y  \\ 
\binclone~\cite{binclone}               &      90 &      18 &      72 &       0 &      0 & \N & \N & \Y & \N  & \N & \N & \N &     \N &    \Y &    1 &  \Y  \\   
\ruttenberg~\cite{Ruttenberg2014Shared} &     936 &       0 &     936 &      -- &      0 & \N & \N & \Y & \N  & \N & \N & \N &     \N &    \Y &    0 &  \Y  \\  
\cesare~\cite{Cesare2014Control}        &  16,999 &   1,601 &  25,398 &       0 &      0 & \N & \N & \Y & \N  & \N & \N & \N &     \Y &    \Y &    0 &  \Y  \\
\blex~\cite{blex}                       &   1,140 &   1,140 &       0 & 0.196 M &      0 & \Y & \Y & \N & \Y  & \Y & \Y & \N &     \N &    \Y &    1 &  \Y  \\ 
\cop~\cite{cop}                         &     321 &     321 &       0 &       0 &      0 & \Y & \Y & \N & \N  & \Y & \Y & \N &     \Y &    \Y &    4 &  \N  \\ 
\tedem~\cite{tedem}                     &      15 &      15 &       0 &      -- &      0 & -- & \N & \Y & --  & \N & \N & \N &     \N &    \Y &    0 &  \Y  \\ 
\sigmasaed~\cite{sigma}                 &      18 &      16 &       2 &      -- &      0 & \N & \N & \Y & \N  & \N & \N & \N &     \N &    \N &    0 &  \N  \\ 
\ming~\cite{Ming2015Memoized}           &     155 &       0 &     155 &       0 &      0 & \N & \N & \Y & \N  & \N & \N & \N &     \N &    \N &    0 &  \Y  \\
\multimh~\cite{Pewny2015BugSearch}      &      60 &      60 &       0 &      -- &      2 & \Y & \N & \N & \Y  & \Y & \Y & \Y &     \N &    \Y &    0 &  \Y  \\
\edg~\cite{edg2015}                     &     384 &     384 &       0 & 0.020 M &      0 & \N & \N & \Y & \N  & \N & \N & \N &     \N &    \Y &    1 &  \Y  \\   
\discovre~\cite{discovre}               &   2,280 &   2,280 &       0 & 0.564 M &      2 & \Y & \Y & \Y & \Y  & \Y & \Y & \Y &     \N &    \Y &    2 &  \Y  \\ 
\mockingbird~\cite{mockingbird}         &      10 &      10 &       0 &      -- &      0 & \Y & \N & \N & \Y  & \Y & \Y & \Y &     \N &    \Y &    2 &  \Y  \\ 
\esh~\cite{esh}                         &   1,000 &   1,000 &       0 &      -- &      0 & \Y & \Y & \N & \Y  & \N & \Y & \N &     \N &    \Y &    2 &  \Y  \\ 
\tpm~\cite{tpm}                         &       7 &       0 &       7 &       0 &      0 & \N & \N & \Y & \N  & \N & \N & \N &     \Y &    \Y &    4 &  \N  \\ 
\bindnn~\cite{bindnn}                   &      -- &   2,068 &       0 & 0.013 M &      0 & \Y & \Y & \N & \N  & \Y & \Y & \Y &     \N &    \Y &    1 &  \Y  \\ 
\genius~\cite{genius}                   &      -- &  17,626 &       0 &   420 M &  8,128 & \Y & \N & \N & \Y  & \Y & \Y & \Y &     \N &    \Y &    3 &  \Y  \\ 
\bingo~\cite{bingo}                     &     110 &     110 &       0 & 0.127 M &      0 & \Y & \N & \N & \Y  & \Y & \Y & \Y &     \N &    \Y &    4 &  \Y  \\ 
\kim~\cite{Kim2016}                     &     350 &      30 &     320 &       0 &      0 & \N & \N & \Y & \N  & \N & \N & \N &     \N &    \N &    1 &  \Y  \\
\kamino~\cite{kam1n0}                   &      10 &      10 &       0 &      -- &      0 &  --& -- & -- & --  & \N & \N & \N &     \N &    \N &    4 &  \Y  \\ 
\binsequence~\cite{binsequence}         &      19 &      17 &       2 &   3.2 M &      0 & \N & \N & \Y & \N  & \N & \N & \N &     \N &    \Y &    4 &  \Y  \\ 
\xmatch~\cite{xmatch}                   &      72 &      72 &       0 & 0.007 M &      1 & \Y & \N & \N & \Y  & \N & \Y & \Y &     \N &    \Y &    4 &  \Y  \\ 
\cacompare~\cite{cacompare}             &      72 &      72 &       0 &      -- &      0 & \Y & \N & \N & \Y  & \Y & \Y & \Y &     \N &    \Y &    3 &  \Y  \\ 
\spain~\cite{spain}                     &      28 &      28 &       0 &      -- &      0 & -- & -- & -- & --  & \N & \N & \N &     \N &    \Y &    0 &  \Y  \\ 
\binsign~\cite{binsign}                 &       9 &       7 &       2 & 0.023 M &      0 & \N & \N & \Y & \N  & \Y & \N & \N &     \Y &    \Y &    2 &  \Y  \\ 
\gitz~\cite{gitz}                       &      -- &      -- &       0 &   0.5 M &      0 & \Y & \Y & \N & \Y  & \N & \Y & \Y &     \N &    \Y &    0 &  \Y  \\ 
\binshape~\cite{binshape}               &      51 &      50 &       1 &     3 M &      0 & \Y & \N & \Y & \N  & \Y & \Y & \N &     \Y &    \Y &    0 &  \Y  \\ 
\binsim~\cite{binsim}                   &   1,062 &       4 &   1,058 &       0 &      0 & \N & \N & \Y & \N  & \N & \N & \N &     \Y &    \Y &    6 &  \Y  \\ 
\kargen~\cite{Kargen2017}               &      11 &      11 &       0 &       0 &      0 & \Y & \N & \N & \Y  & \Y & \Y & \N &     \Y &    \Y &    0 &  \Y  \\ 
\imfsim~\cite{imfsim}                   &   1,140 &   1,140 &       0 &      -- &      0 & \Y & \Y & \N & \Y  & \Y & \Y & \N &     \Y &    \Y &    3 &  \Y  \\ 
\Gemini~\cite{gemini}                   &  51,314 &  51,314 &       0 &   420 M &  8,126 & \Y & \N & \N & \N  & \Y & \N & \Y &     \N &    \Y &    2 &  \Y  \\ 
\fossil~\cite{fossil}                   &   6,925 &   1,920 &   5,005 &   1.5 M &      0 & \Y & \Y & \Y & \Y  & \N & \N & \N &     \Y &    \Y &    7 &  \Y  \\
\firmup~\cite{firmup}                   & 200,000 & 200,000 &       0 &    40 M &  2,000 & -- & -- & -- & --  & \N & \N & \Y &     \N &    \Y &    2 &  \Y  \\ 
\binarm~\cite{binarm}                   &       0 &       0 &       0 &   3.2 M &  5,756 & \Y & \N & \N & \N  & \Y & \N & \N &     \N &    \Y &    5 &  \Y  \\
\alphadiff~\cite{alphadiff}             &  67,427 &  67,427 &       0 &       0 &      2 & \Y & \N & \N & \Y  & \Y & \Y & \Y &     \N &    \Y &    6 &  \N  \\
\vulseeker~\cite{vulseeker}             & --      &      -- &       0 & 0.736 M &  4,643 & \Y & \N & \N & \N  & \Y & \N & \Y &     \N &    \Y &    1 &  \Y  \\
\jointlearning~\cite{jointlearning}     &      -- &     --  &       0 & 0.202 M &      0 & \N & \N & \N & \Y  & \Y & \N & \Y &     \N &    \Y &    0 &  \Y  \\
\innereye~\cite{innereye}               &      -- &     --  &       0 &   1.2 M*&      0 & \N & \N & \N & \Y  & \Y & \N & \Y &     \N &    \Y &    0 &  \Y  \\
\asmtovec~\cite{asm2vec}                &   1,116 &   1,116 &       0 & 0.140 M &      0 & \Y & \Y & \N & \Y  & \Y & \Y & \N &     \Y &    \Y &   12 &  \Y  \\
\safe~\cite{safe2019}                   &      -- &      -- &       0 & 1.847 M &      0 & \Y & \N & \N & \Y  & \Y & \Y & \Y &     \N &    \Y &    1 &  \Y  \\

\hline
                                                                                                                                           
\end{tabular}                                                                                                                              
}                                                                                                                                          
\end{table*}

\subsection{Datasets}
\label{sec:datasets}

The left side of Table~\ref{tab:evaluation} describes the datasets used 
by each approach. 
It first shows 
the total number of executables used in the evaluation and 
their split into benign and malicious executables. 
Executables may come from different programs or correspond to
multiple versions of the same program, 
e.g., with varying compiler and compilation options.
Then, for approaches that have function granularity, it captures the total
number of functions evaluated, 
and for approaches that analyze firmware, 
the number of images from where the executables are obtained.
A dash (--) in a column means that we could not find the number in the paper.
For example, \saebjornsen evaluates on system library files in Windows XP,
but the total number of executables is not indicated.

Most approaches use custom datasets, 
the one popular benchmark is Coreutils used by 18 approaches. 
In addition, approaches that evaluate on firmware use 
two openly available firmwares 
(ReadyNAS~\cite{readynas} and DD-WRT~\cite{ddwrt}).
Over half of the approaches evaluate on 
less than 100 executables, 
7 on less than 1K, 
8 on less than 10K, and
only 8 on over 10K.
Out of \numapproaches approaches,
37 have evaluated only on benign programs,
8 only on malware, and 16 on both.
This indicates that \bcs is also popular for malware analysis.
However, of the 24 approaches evaluated on malware,
only five use packed malware samples 
(\smit, \beagle, \mutantxs, \cesare, \binsim). 
These approaches first unpack the malware using a 
custom unpacker (\smit) or a generic (write-and-execute) unpacker 
(\beagle, \mutantxs, \cesare, \binsim), and then compute \bcs.
The rest have access to the malware's source code or to 
unpacked samples. 

For \search approaches that use function granularity, 
the number of functions in the repository better captures the dataset size. 
The largest dataset is by \genius, 
which evaluates on 420M functions extracted from 8,126 firmwares.
Prior approaches had evaluated on at most 0.5M functions, 
which demonstrates the scalability gains from its embedding approach. 
Five other approaches have evaluated on over 1M functions:
\fossil (1.5M), 
\binsequence (3.2M), 
\binarm (3.2M), 
\firmup (40M), and
\Gemini (420M), which uses the same dataset as \genius.
In addition, \innereye has evaluated on a repository of 1.2M basic blocks.

\subsection{Methodology}
\label{sec:evalmethods}

The right side of Table~\ref{tab:evaluation} describes four aspects of 
the evaluation methodology used by each approach:
robustness,
accuracy,
performance,
and comparison with prior approaches.

\paragraph{Robustness.}
The first 8 columns on the right side of Table~\ref{tab:evaluation} 
capture how authors evaluate the robustness of their approaches,
i.e., their ability to capture similarity despite transformations 
applied to the input programs.
First, it shows whether they use each of the four compilers we have 
observed being employed to compile the programs in the dataset: 
GCC, ICC, Visual Studio (MSVS), and Clang.
Then, it captures whether the authors evaluate similarity between 
programs compiled with different compilation options (cross-optimization), 
between programs compiled with different compilers (cross-compiler), and 
between programs compiled for different architectures (cross-architecture). 
Finally, it captures whether the authors evaluate similarity when 
obfuscation transformations are applied to the input programs.

There are 34 approaches that evaluate robustness 
(at least one \Y in the last four robustness columns) and 27 that do not.
Many early works did not evaluate robustness. 
This evaluation has become increasingly popular as approaches mature.
The most popular robustness evaluation is 
cross-optimization (23 approaches), followed by 
cross-compiler (19), cross-architecture (\numcrossarch), and obfuscation (13).
There are 9 approaches that have evaluated cross-optimization, cross-compiler, and 
cross-architecture.
Approaches that evaluate cross-compiler also typically evaluate 
cross-optimization, as it is a simpler case.
Simlarly, approaches that evaluate cross-architecture 
typically also evaluate cross-compiler, 
as cross-architecture programs may be produced using different compilers.
Note that it is possible for approaches to compile programs with multiple 
compilers, but not perform cross-compiler evaluation, 
i.e., not compare similarity between programs compiled with different 
compilers.

There are 13 approaches that have evaluated on obfuscated programs.
Of those, two use only source code transformations (\cop, \asmtovec),
three use only binary code transformations (\kruegel, \tpm, \binshape),  
five use packed malware (\smit, \beagle, \mutantxs, \cesare, \binsim), and
three evaluate both source code and binary code transformations 
(\binsim, \imfsim, \fossil).

\paragraph{Accuracy evaluation and comparison.}
There are 49 approaches that perform a quantitative evaluation 
of their accuracy using some ground truth (\Y), and 
12 that perform qualitative accuracy evaluation through case studies (\N). 
Quantitative evaluation most often uses standard accuracy metrics such as 
true positives, false positives, precision, and recall. 
However, two approaches propose novel application-specific accuracy 
metrics (\iline, \kargen).

There are 33 approaches that compare with prior approaches.
All of them compare accuracy and six also compare runtime.
The top target for comparison is \bindiff (13 approaches compare with it),
followed by \tracy (5), \discovre (4), and \multimh (4).
Comparing accuracy across \bcs approaches is challenging for 
multiple reasons. 
First, only a small fraction of the proposed approaches have publicly released 
their code (Section~\ref{sec:implementation}). 
Since most approaches are not publicly available, 
comparison is often performed by re-implementing previous approaches, 
which may require significant effort.
One advantage of reimplementation is that approaches can be compared
on new datasets.
The alternative to re-implementation is to evaluate the new approach on 
the same dataset used by a prior approach, 
and compare with the reported results. 
This method is only used by 6 approaches
(\genius, \bingo, \xmatch, \cacompare, \Gemini, and \binarm) 
likely because most datasets are custom and not publicly available. 
Fortunately, we observe that public code release has become more 
common in recent approaches.
Second, the input comparison and input granularity may differ among approaches 
making it nearly impossible to perform a fair comparison. 
For instance, it is hard to compare in a fair manner an approach that 
identifies program similarity using callgraphs (e.g., \smit) 
with an approach comparing basic blocks (e.g., \innereye).
Third, even when the input comparison and input granularity are the same, 
the evaluation metrics and methodology may differ,
significantly impacting the measured accuracy. 

The latter challenge is best illustrated on \search approaches operating at 
function granularity.
These approaches find the most similar functions in a repository to a 
given function.
They return multiple entries ranked in descending similarity order and
count a true positive if one of the \emph{top-k} most 
similar entries is a true match. 
Unfortunately, the values of $k$ vary across approaches and significantly 
impact the accuracy, 
e.g., a 98\% precision on top 10 is significantly worse than a 
98\% precision on top 3.
Thus, it becomes hard to compare accuracy numbers obtained with different 
$k$ values and it becomes tempting to raise $k$ until a sufficiently 
high accuracy number is achieved.
Furthermore, many approaches do not describe the similarity threshold 
used to determine that no similar entry exists in the repository. 
This means that they always find some similar entry in the repository, 
even if the similarity may be really low.

\paragraph{Performance.}
It is common (49/\numapproaches) to measure the runtime performance of an approach.
Runtime is typically measured end-to-end, 
but a few approaches report it for each approach component (e.g., \binsim).
Four approaches report their asymptotic complexity
(\bruschi, \saebjornsen, \iline, \mockingbird).

\section{Discussion}
\label{sec:discussion}

This section discusses open challenges and possible future research directions.

\paragraph{Small \chunks.}
Many \bcs approaches ignore small \chunks, 
setting a threshold on the minimum number of instructions or basic blocks 
to be considered. 
Oftentimes, only \chunks with a handful of instructions are ignored, 
e.g., functions with less than 5 instructions, 
but some approaches use large thresholds like 100 basic blocks in \tracy. 
Small \chunks are challenging because they are common, 
may comprise of a single basic block that prevents structural analysis, and 
may have identical syntax despite different semantics. 
For example, setter functions that update the value of a field in an object
have nearly identical syntax and structure, 
simply setting a memory variable with the value of a parameter.
But, they may have very different semantics,
e.g., setting a security level or updating a performance counter, 
Furthermore, semantic similarity techniques like instruction classification, 
symbolic formulas, and input-output pairs may fail to capture their 
differences, e.g., if they do not distinguish different memory variables. 
Similarity of small \chunks remains an open challenge. 
One potential avenue would be to further incorporate context. 
Some structural approaches already consider the callgraph to match functions 
(e.g., \flake), but this does not always suffice.
We believe that it may be possible to further incorporate other context 
like locality (e.g., how close the binary code is in the program structure) or 
data references (e.g., whether they use equivalent variables).

\paragraph{Source-to-binary similarity.}
Some applications like plagiarism detection 
may require source code to be compared with binary code. 
Early approaches for source-to-binary similarity used software birthmarks
that capture inherent functionality of the
source code~\cite{Myles2005KBS,Choi2007SBB}.
Recently, source-to-binary similarity has been applied for 
searching if a known bug in open source code exists in 
some target binary code~\cite{fiber}.
The availability of source code provides more semantics 
(e.g., variable and function names, types, comments) 
potentially improving accuracy compared to simply  
compiling the source code and performing \bcs.
We believe other applications remain that require 
determining if a target \chunk has been compiled, or has evolved, 
from some given source code. 
For example, there may be programs for which source code is only available 
for an old version and there is a need to understand how newer binary versions 
have evolved from the original source code.  

\paragraph{Data similarity.}
This survey has focused on \bcs, 
but programs comprise both code and data. 
There may be situations where the data used by the code is as important as 
the code, e.g., when the only change between versions of a program is in the 
data used such as changing parameters for a machine learning classifier.
Furthermore, data may be stored in complex data structures that may be 
key to the functionality. 
There exists a long history of type inference techniques on 
binary code~\cite{types}, 
which we believe could be combined with \bcs to compare the data structures 
used by different \chunks, or how two \chunks use the same data structure.

\paragraph{Semantic relationships.}
A different aspect of semantic similarity is to identify binary code 
with \emph{related} functionality, 
e.g., cryptographic or networking functions.
This is challenging as code that is related in its functionality 
may not have the same semantics. 
For example, a decryption function is clearly related to its 
encryption function, but performs opposite operations.
So far, most work has focused on domain specific techniques such as those 
for identifying cryptographic functions 
(e.g.,~\cite{wang2009reformat,xu2017cryptographic}). 
But, recently some approaches have started exploring domain-agnostic 
techniques~\cite{bcd}.
We believe further work is needed to better define such semantic relationships 
and their identification.

\paragraph{Scalability.}
The largest \bcs evaluation so far is by \genius and \Gemini on 420M functions 
from 8,126 firmware images (500K functions per image).
While that is a significant step from prior approaches, 
if we consider instead 100K unique firmware, 
a conservative number since it is expected that there will be 
20 Billion IoT devices connected to the Internet by 2020~\cite{gartnerIot},
we need a \bcs approach that can handle 50 Billion functions. 
Thus, further improvements on scalability will be needed to realize the 
vision of binary code search engines.

\paragraph{Obfuscation.}
Many challenges still remain for \bcs on obfuscated code. 
For example, a recent study has shown that state of the art 
unpacking techniques can miss 20\%--60\% of the original code, 
e.g., due to incomplete function detection~\cite{lineage}.
And, no \bcs approach currently handles virtualization-based packers 
such as Themida~\cite{themida} and VMProtect~\cite{vmprotect},
which generate a random bytecode instruction set, 
transform an input \chunk into that bytecode instruction set, and 
attach an interpreter (or virtual machine) 
from the generated bytecode to native code.
Furthermore, obfuscation is best addressed with semantic similarity, 
which has a challenge with obfuscation transformations that do not 
respect the original semantics of the code, but still perform its main goals.

\paragraph{Approach comparison.}
The variety of datasets and methodologies used to evaluate the approaches 
(e.g., top $k$ evaluation for OM approaches), 
together with the absence of source code for many of them, 
makes it hard to perform a fair comparison to understand 
their benefits and limitations. 
We believe the field would greatly benefit from open datasets for benchmarking, 
and independent validation under the same experimental conditions.
Furthermore, there is a need to improve the comparison of approaches 
that handle obfuscation, beyond changes of compiler and compiler options. 
Building a dataset that covers real-world obfuscations is fundamental 
given the huge space of possible obfuscation transformations and that
each approach supports a different subset.

\section{Conclusion}
\label{sec:conclusion}

During the last 20 years, 
researchers have proposed many approaches to perform \bcs and 
have applied them to address important problems such as 
patch analysis, bug search, and malware detection and analysis.
The field has evolved  
from \diffing to \search;
from syntactic similarity to incorporate structural and semantic similarity; 
to cover multiple code granularities,
to strengthen the robustness of the comparison
(i.e., cross-optimization, cross-compiler, cross-OS, cross-architecture, 
obfuscation); 
to scale the number of inputs compared 
(e.g., through hashing, embedding, indexing); and 
to automatically learn similarity features.

Despite its popularity, the area had not yet been systematically analyzed.
This paper has presented a first survey of \bcs.
The core of the paper has systematized \numapproaches \bcs 
approaches on four dimensions:
the applications they enable, 
their approach characteristics, 
how the approaches are implemented, and 
the benchmarks and evaluation methodologies used to evaluate them.
It has discussed the advantages and limitations of 
different approaches, their implementation, and their evaluation. 
It has summarized the results into easy to access tables useful 
for both beginners and experts. 
Furthermore, it has discussed the evolution of the area and 
outlined the challenges that remain open, 
showing that the area has a bright future.

\paragraph{Acknowledgments.}
This research was partially supported
by the Regional Government of Madrid through the
BLOQUES-CM (S2018/TCS-4339) grant; and
by the Spanish Government through the 
SCUM (RTI2018-102043-B-I00) grant.
This research is also supported by the EIE Funds of the European Union.
All opinions, findings and conclusions, or recommendations expressed herein
are those of the authors and do not necessarily reflect the views
of the sponsors.

{

}

\section*{Appendix}

\begin{table} [h]
\centering
\tiny
\caption{Comparison of the binary analysis platforms used in the implementation of \bcs approaches.
\textbf{This table overlaps with previous work on type inference~\cite{types}.}
We have added following platforms: Angr, Beaengine, Cuckoo, NewBasic, McSema, Vulcan.
\label{tab:platforms}}
{
\setlength\tabcolsep{4pt}
\renewcommand\featuretext[1]{\rotatebox{90}{#1}}
\begin{tabular}{|r|l|ll||ll|llll|ll||lll|}
\cline{5-15}
\multicolumn{4}{c|}{} & \multicolumn{8}{c||}{\textbf{Target Arch. \& OS}} & \multicolumn{3}{c|}{\textbf{Release}} \\
\hline
& & & & \multicolumn{2}{c|}{\textbf{CISC}} & \multicolumn{4}{c|}{\textbf{RISC}} &  \multicolumn{2}{c||}{\textbf{OS}} & \multicolumn{3}{c|}{} \\
{\textbf{Platform}} & {\textbf{IR}} & \featuretext{\textbf{Static analysis}} & \featuretext{\textbf{Dynamic analysis}} & \featuretext{\textbf{x86}} & \featuretext{\textbf{x86-64}} & \featuretext{\textbf{ARM}} &\featuretext{\textbf{MIPS}}  &   \featuretext{\textbf{SPARC}}  &   \featuretext{\textbf{Power PC}}  &    \featuretext{\textbf{Windows / PE}}  & \featuretext{\textbf{Linux / ELF}} &    \featuretext{\textbf{Open source}} &    \featuretext{\textbf{Free binary}} &    \featuretext{\textbf{Commercial}} \\
\hline
Angr~\cite{angrUrl}         & VEX       & \Y & \N & \Y & \N & \Y & \Y & \N & \N & \Y & \Y & \Y & \N & \N \\
BAP~\cite{BAP}              & BIL       & \Y & \Y & \Y & \Y & \Y & \N & \N & \N & \Y & \Y & \Y & \N & \N \\
Beaengine~\cite{beaengine}  & \N        & \Y & \N & \Y & \N & \N & \N & \N & \N & \Y & \N & \Y & \N & \N \\
BitBlaze~\cite{bitblazeUrl} & VINE      & \Y & \Y & \Y & \N & \N & \N & \N & \N & \Y & \Y & \Y & \N & \N \\
Boomerang~\cite{Boomerang}  & RTL       & \Y & \N & \Y & \N & \N & \Y & \Y & \Y & \Y & \Y & \Y & \N & \N \\  
Cuckoo~\cite{cuckoo}        & \N        & \N & \Y & \Y & \Y & \Y & \N & \N & \N & \Y & \Y & \Y & \N & \N \\
Dyninst~\cite{dyninstUrl}   & \N        & \Y & \Y & \Y & \Y & \N & \N & \N & \N & \N & \Y & \Y & \N & \N \\
IDA~\cite{ida}              & IDA       & \Y & \N & \Y & \Y & \Y & \Y & \Y & \Y & \Y & \Y & \N & \N & \Y \\
LLVM~\cite{llvmUrl}         & LLVM-IR   & \Y & \N & \N & \N & \N & \N & \N & \N & \N & \N & \Y & \N & \N \\
NewBasic~\cite{newbasicUrl} & \N        & \Y & \N & \Y & \N & \N & \N & \N & \N & \N & \N & \Y & \N & \N \\
McSema~\cite{McSema}        & LLVM-IR   & \Y & \N & \Y & \Y & \Y & \N & \N & \N & \Y & \Y & \Y & \N & \N \\
PIN~\cite{pinUrl}           & \N        & \N & \Y & \Y & \Y & \N & \N & \N & \N & \Y & \Y & \N & \Y & \N \\
QEMU~\cite{QEMU}            & TCG       & \N & \Y & \Y & \Y & \Y & \Y & \Y & \Y & \Y & \Y & \Y & \N & \N \\
ROSE~\cite{ROSE}            & SAGE-III  & \Y & \N & \Y & \N & \Y & \Y & \Y & \Y & \Y & \Y & \Y & \N & \N \\
Valgrind~\cite{valgrindUrl} & VEX       & \N & \Y & \Y & \Y & \Y & \Y & \Y & \Y & \Y & \Y & \Y & \N & \N \\
Vulcan~\cite{vulcan}        & VULCAN-IR & \Y & \N & \Y & \Y & \N & \N & \N & \N & \Y & \N & \N & \N & \N \\
\hline
\end{tabular}
}
\end{table}

\paragraph{Implementation platforms.}
Table~\ref{tab:platforms} shows
the intermediate representation (IR) used by the platform,
whether it supports static and dynamic analysis,
the target architectures and operating systems it supports, and
how it is released (open source or free binary).
Among the \numplatforms platforms, 12 support static analysis and
7 dynamic analysis.
The functionality provided by static analysis platforms widely varies.
\ida is a disassembler,
Boomerang is a decompiler, and the rest
offer diverse static analysis functionality such as disassembly,
building control flow graphs and call graphs,
IR simplifications, and data flow propagation.
All dynamic analysis platforms can run unmodified binaries
(i.e., no need to recompile or relink).
QEMU is a whole system emulator that can run a full guest OS (e.g., Windows)
on a potentially different host OS (e.g., Linux).
Dyninst, PIN, and Valgrind execute an unmodified target program with
customizable instrumentation (e.g., through binary rewriting)
on the CPU of the host system.
BAP and \bitblaze
build their dynamic analysis components on top of PIN and QEMU.

%\clearpage

\end{document}